\documentclass[journal]{IEEEtran}

\usepackage{cite}
\usepackage{t1enc}
\usepackage[cmex10]{amsmath}
\usepackage{amssymb}
\usepackage{array}
\usepackage{mdwmath}
\usepackage{mdwtab}
\usepackage[dvipsnames]{xcolor}
\usepackage{pgfplots}

\usetikzlibrary{plotmarks}
\pgfplotsset{compat=newest}
\pgfplotsset{plot coordinates/math parser=false}
\newlength\figureheight
\newlength\figurewidth 
\PassOptionsToPackage{pdf}{pstricks,pst-node,pst-plot}
\usepackage{pstricks,pst-node,pst-plot}
\usepackage{graphicx}
\usepackage{amsmath,amssymb,array,multirow,bigdelim,dsfont}
\usepackage[amssymb]{SIunits} 
\usepackage{url}
\usepackage{t1enc}
\usepackage{url}   
\usepackage{booktabs}
\usepackage{array}
\usepackage[draft]{fixme}

\begin{document}
\title{Stochastic Multipath Model for the In-Room Radio Channel based on Room
  Electromagnetics}
\author{Troels~Pedersen\thanks{\today,  This work is supported 
by  the  Cooperative  Research  Project
VIRTUOSO,  funded  by  Intel  Mobile
Communications, Keysight, Telenor, Aalborg University, and the Danish National
Advanced Technology Foundation. This work was performed within
the framework of the COST Action CA15104 IRACON.
T.  Pedersen  is  with  the  Department   of  Electronic
Systems,  Section  Wireless  Communication  Networks,  Aalborg  University,
Aalborg, 9220, Denmark (e-mail:  troels@es.aau.dk). } }

\maketitle

 \begin{abstract}
   We propose a stochastic multipath model for the received signal for the case
   where the transmitter and receiver, both with directive antennas, are situated
   in the same rectangular room.    This scenario is known to produce
   channel impulse responses with a gradual specular-to-diffuse
   transition in delay.  Mirror source theory predicts the arrival rate to
   be quadratic in delay, inversely proportional to room volume and
   proportional to the product of the antenna beam coverage fractions. We
   approximate the mirror source positions by a homogeneous spatial
   Poisson point process and their gain as complex random variables with
   the same second moment.  The multipath delays in the resulting model
   form  an inhomogeneous Poisson  point process which enables
   derivation of  the characteristic  functional, power/kurtosis delay
   spectra, and the distribution of order statistics of the arrival
   delays in closed form. We find that the
   proposed model matches the  mirror source model well in terms of
   power delay spectrum, kurtosis delay spectrum, order statistics,
   and prediction of mean delay and rms delay spread.
 The constant rate model, assumed in e.g. the Saleh-Valenzuela model,
 is unable to reproduce the same effects.
 \end{abstract}


\section{Introduction}
\label{sec:introduction}
\PARstart{S}{tochastic} models for 
multipath channels are useful tools for the design, analysis and
simulation of systems for radio localization and communications. These models allow for tests via
Monte Carlo simulation and in many cases provide analytical results
useful for system design. Numerous such models exist for the complex
baseband representation of the signal at the receiver
antenna  (omitting any additive terms due to noise or
interference) 
\begin{equation}
  \label{eq:66}
  y(\tau) = \sum_{k }\alpha_k s(\tau-\tau_k),
\end{equation}
where $s(t)$ is the complex baseband representation of the transmitted
signal and the term due to path~$k$ has complex gain $\alpha_k$ and delay $\tau_k$.
The received signal is fully  described as a marked point process 
\begin{equation}
  \label{eq:28}
\mathcal X =\{(\tau_0,\alpha_0), (\tau_1,\alpha_1),(\tau_2,\alpha_2),\dots\}.
\end{equation}
A models of this structure was studied in the pioneering work by Turin \cite{turin} in which $\mathcal X$ can be seen as a marked Poisson
point process specified by parameters determining the arrival rate
$\lambda(\tau)$ and the mark density $p(\alpha|\tau)$. 
Although Turin's model was originally intended for urban radio
channels, it has since
been taken as a the basis for a wide range of models for outdoor and
indoor channels including the clustered  models by Suzuki~\cite{suzuki},  Hashemi~\cite{hashemi},
Saleh and Valenzuela \cite{saleh}, Spencer et al. \cite{Spencer2000} and
Zwick et al. \cite{Zwick2002,Zwick2000}. More recently, this type of statistical
channel models  has been considered for ultrawideband 
\cite{Chong2005, Molisch2006} and for millimeter-wave spectrum
\cite{Gustafson2014 ,Haneda2015,Samimi2016} systems. 
To make use of the model, the arrival rate and mark density should be defined.
These settings are critical  since they determine channel parameters relavant for 
system design such as the distribution of instantaneous mean delay and rms delay spread.

Commonly, the arrival rate (within a cluster) is assumed constant while the second moment of the
mark density  is assumed to follow an exponential decay
\cite{saleh,Spencer2000,Zwick2002,Zwick2000, Gustafson2014, Chong2005,
  Molisch2006,Haneda2015,Samimi2016}.  The ``constant rate'' model is
appealing since it requires only one
single parameter, i.e. the arrival rate,  to be determined
empirically.  Usually a two-step procedure is followed: 
first the points of the hidden process $\mathcal X$ are estimated from 
observations of $y(t)$ and then the arrival rate is estimated from there
(e.g. relying on  interarrival times). The first step of
this method is prone to censoring effects as discussed in \cite{Karttunen2017}. 
In the presense of noise, the weaker components predominantly arriving
at large delays may be undetected. A similar effect occurs if the 
measurement bandwidth is insufficient to distinguish signal 
components with short interarrival times. If unaccounted for, both of
these censoring effects lead to  underestimation of the arrival rate. 
As noted in \cite{Meijerink2014}, several authors  justify the
constant rate  assumption  qualitatively as a ``convenient compromise'' 
between the increasing number of possible multipath components and the
increasing shadowing probability.  Nevertheless, as also noted in
\cite{Meijerink2014}, there seems to be no principal reason that the
effects should balance each other out to produce exactly a constant
rate.  

In some cases, stochastic multipath models relying on the constant
rate model do not agree well with measurements at all. This is
particularly true for inroom propagation channel,  which has been explored in a  number of works
including
\cite{Holloway1999,Kunisch2003,Kunisch2002,Steinboeck2016,Steinboeck2013d,Pedersen2006,Pedersen2007,Pedersen2012,Pedersen2018}.
There the received signal
exhibits a  gradual diffusion from specular at
early delays to diffuse at later delays. This effect is  not captured  in the
constant rate model.  Moreover, in clustered models, the
cluster arrival rate is assumed constant yielding a total 
path arrival rate which increases linearly with delay
\cite{Jakobsen2012}. This increase is, however, too slow compared to
the power decay to account for  the specular-to-diffuse transition.




For the inroom scenario, an alternative to the constant rate assumption was proposed in
\cite{Pedersen2018}. The model is based on mirror source analysis of
the case where the transmitter and receiver antennas are both
directive and sit within the same  rectangular room with flat
walls. For this setup, the arrival rate (averaged over uniformly
distributed transmitter antenna positions and orientations) was found
to be inversely proportional to room volume, depend on antenna
directivity and to increase quadratically with delay accounting for
the specular-diffuse transition.  The analysis also leads to closed
form expressions for the variance of the mark density and the
resulting power delay spectrum. The power delay spectrum decays
exponentially in delay and the reverberation time (decay rate) is
predicted very accurately by the Eyring model
\cite{Eyring1930,Holloway1999,Steinbock2015} applying Kuttruff's
correction factor \cite{Kuttruff2000}. The power delay spectrum is of
the same form as studied and experimentally validated in
\cite{Steinbock2015,Steinboeck2013}. The analysis in
\cite{Pedersen2018} further revealed that while the arrival rate  was easy
to derive, higher order moments for the arrival process are very
difficult to obtain from the mirror source theory.


In the present contribution, we propose to approximate the mirror
source model in \cite{Pedersen2018} using a  spatial Poisson process
which is more analytically tractable.  The obtained model is of the same type as
Turin's model, but has the same arrival rate and power delay spectrum
as the mirror source process. The approximation model permits
derivation of  moments and cumulants of the received signal via the
characteristic and cumulant generating functionals. Thus, we derive  
the power (second moment) and kurtosis (ratio of fourth
moment and squared second moment) of the received signal as a function
of delay.  The kurtosis is related to the arrival rate: in the
simplifying large bandwidth case, the arrival rate is inversely
proportional to the excess kurtosis of the received signal.
For the proposed inhomogeneous model, the distribution of interarrival
times, which has been widely studied in the channel modeling
literature, turns out to be  degenerate and is therefore not useful. 
Instead, we study the distribution of order statistics which we derived
for the approximation model.

To  evaluate the accuracy of the proposed Poisson
approximation we compare it to Monte Carlo simulations of the mirror
source model. Our simulations show that the proposed Poisson
approximation captures the specular-diffuse transition and fit well
the mirror source model in terms of power- and
kurtosis delay spectra and  order statistics for the arrival
process. The distribution of instantaneous mean delay and
rms delay spread agrees with the mirror source model. 
In these simulations we also contrast with the constant
rate model. It is observed that the constant rate model  is able to
capture the power delay  spectrum, but represents poorly the kurtosis,
order statistics, instantaneous mean delay and rms delay spread. We
conclude that to accurately  model these parameters for in-room channels, the constant rate
model is inadequate.

The paper is organized as
follows. Section~\ref{sec:mirrorSourceTheory} defines the notation and
summarizes the results of \cite{Pedersen2018}. In
section~\ref{sec:mirr-source-proc} we first represent the mirror
source process as a spatial point process and then approximate it
using a Poisson point process. Sections~\ref{sec:stat-moments-cumul}
and~\ref{sec:arrivalTimes} gives the results related to kurtosis of
the received signal $y(t)$ and the distribution of order statistics
for the arrival times. The accuracy of the proposed Poisson
approximation is tested in numerical examples given in
Section~\ref{sec:simul-stud-exampl}. Discussion and conclusions are
given in \ref{sec:discussion} and \ref{sec:conclusion}.


\section{Mirror Source Process for Rectangular Room}
\label{sec:mirrorSourceTheory} 
The present contribution relies on the same setup as in the previous work
 \cite{Pedersen2018} and utilizes the results sumarized below.
For further details, the reader is referred to \cite{Pedersen2018}.

Consider a rectangular room with two directive antennas 
(one transmitter and one receiver) located inside. The room has dimension $L_x\times
L_y \times L_z$,  volume $V=L_xL_yL_z$ and surface area $S=2(L_xL_y+L_yL_z+L_xL_z)$.
Positions are given in a Cartesian coordinate system aligned such that the room spans the
set $[0,L_x)\times[0,L_y)\times[0,L_z)$.  We assume  that the carrier
wavelength $l_c$ is small
compared to the room dimensions, and that only specular reflections
occur with an average gain $\bar g$.  The positions of the
transmitter and receiver are denoted by $r_T$ and $r_R$ . We subscript
all entities related to the transmitter and receiver
 by $T$ and $R$, respectively. 

Denote by  $G(\Omega)$  the antenna
 gain in the direction specified by the three dimensional unit vector $ \Omega \in \mathbb
S_2$, where $\mathbb S_2$ is the unit sphere. We assume the antennas
to be lossless. The footprint of an
antenna on the unit sphere surrounding it is  defined as $ \mathcal O
= \{ \Omega: G(\Omega)\geq \epsilon\cdot  G_{\max}\}$ where
$G_{\max}$ is the maximum gain and  $\epsilon \geq 0$ defines a
gain level below which we ignore any signal contributions. The beam
coverage fraction is further defined as  
\begin{equation}
  \label{eq:26}
\omega =   \frac{1}{4\pi}\int_{\mathbb S_2} \mathds 1 (\Omega \in
\mathcal O )  d\Omega,
\end{equation}
where $\mathds 1(\cdot)$ denotes an indicator function with value one
if the argument is true and zero otherwise.  The beam coverage
fraction ranges from zero to one and can be interpreted as the
probability of a wave impinging from a uniformly random direction is
within the antenna beam. 

The mirror sources (and thus the propagation paths) are indexed by a triplet
$k=(k_x,k_y,k_z)$. Mirror source $k$ has position
\begin{equation}
  \label{eq:68}
r_{T(k_x,k_y,k_z)}
 = 
\begin{bmatrix}
\big\lceil \tfrac{k_x}{2} \big\rceil \cdot 2 L_x + (-1)^{k_x}\cdot x_{T}\\
\big\lceil \tfrac{k_y}{2} \big\rceil \cdot 2 L_y + (-1)^{k_y}\cdot y_{T}\\
\big\lceil \tfrac{k_z}{2} \big\rceil \cdot 2 L_z + (-1)^{k_z}\cdot z_{T}  
\end{bmatrix}.
\end{equation}
Further interpretation of the mirror
source index is given in \cite{Pedersen2018}. By
replacing subscript $T$ by subscript $R$ in \eqref{eq:68}, gives the
position  $r_{Rk}$  of  mirror receiver  $k$. Propagation path~$k$  has delay
\begin{equation}
  \label{eq:8}
 \tau_k = \|  r_{Tk} -  r_R\|/c = \|  r_{Rk} -  r_T\|/c, 
\end{equation}
where $c$ is the speed of light. For path~$k$  the direction of
arrival reads 
\begin{equation}
  \label{eq:41}
 \Omega_{Rk} = \frac{ r_{Tk}- r_R}{\| r_{Tk}- r_R\|}.
\end{equation}
The direction of departure denoted by $\Omega_{Tk}$ follows from
\eqref{eq:41} by interchanging subscripts $T$ and $R$. The power gain
of path $k$ is specified as \footnote{Here we consider the special case of all walls
  having the same gain value. The gain for the  more general case with different
  wall gains is stated in \cite{Pedersen2018}.}
\begin{align}
  \label{eq:147}
  |\alpha_{k}|^2 &= \bar g^{|k|} \cdot \frac{G_T(\Omega_{Tk})
    G_R(\Omega_{Rk})}{(4\pi c\tau_k / l_c)^2} 
\end{align}
with the convention $|k| = |k_x|+|k_y|+|k_z|$.

Randomness is introduced to the mirror source model by letting the
transmitter's position be  independent and uniformly distributed
random variables. The arrival count $N(\tau)$, is a random counting variable
designating the number of received (non-zero) signal components with
delay less than or equal to 
$\tau$. The mean arrival count reads 
%
%
 \begin{align}
&\mathbb E[N(\tau)]=\frac{4\pi c^3\tau^3}{3V} \omega_T \omega_R \mathds 1 (\tau>0)
\label{eq:75}
\end{align}
with the corresponding arrival 
rate  
\begin{equation}
  \label{eq:63}
  \lambda(\tau) = \frac{d \mathbb E[N(\tau)]}{d\tau} = \frac{4\pi c^3\tau^2}{V} \omega_T \omega_R \mathds 1 (\tau>0).
\end{equation}

Assuming the complex gains to  be uncorrelated  random variables, the
second moment of the received signal can be written in terms of the
delay power spectrum $P(t)$ as
\begin{equation}
  \label{eq:93}
  \mathds E [|y(\tau)|^2 ] = \int_{-\infty}^{\infty} P(\tau-t) |s(t)|^2 dt. 
\end{equation}
The delay power spectrum can be obtained as a product of the arrival
rate $\lambda(\tau)$ and the conditional second moment of the complex gains
$\sigma_\alpha^2(\tau) = \mathbb E[|\alpha|^2|\tau]$, i.e.
\begin{equation}
  \label{eq:67}
P(\tau)  = \sigma_\alpha^2(\tau)
\lambda(\tau) .
\end{equation}
With close approximation, the conditional second moment is
 \begin{align}
   \label{eq:12}
 \sigma_\alpha^2(\tau) &= \frac{1}{ c^2\tau^2 \omega_T\omega_R}\cdot \exp(-\tau/T). 
 \end{align}
with the reverberation time  defined according to   Eyring's model~\cite{Eyring1930}
\begin{equation}
  \label{eq:103}
  T =  -\frac{4V \xi}{cS\ln(\bar g)},
\end{equation}
where Kuttruff's correction factor~\cite{Kuttruff2000} is
\begin{equation}
  \label{eq:1}
\xi = \frac{1}{1+\gamma^2\ln(g)/2}. 
\end{equation}
Here, the constant $\gamma^2$ depends on the aspect ratio of the
room and can be found from Monte Carlo simulations. It typically ranges from  0.3 to 0.4 \cite{Kuttruff2000}. The resulting power delay spectrum reads
\begin{equation}
  \label{eq:84}
  P(\tau) = 
 \mathds 1 (\tau>0) \cdot 
  \frac{4\pi c 
}{ V} \cdot \exp(-\tau/T).  
\end{equation}
Notice that the antennas do not enter in \eqref{eq:84}.

\section{The Mirror Source Process as a Spatial Point Process}
\label{sec:mirr-source-proc}
The mirror source model can be studied by viewing the positions of the mirror sources
\begin{equation}
  \label{eq:6}
\mathcal M = \{r_{Tk} : k\in\mathbb Z^3\}.  
\end{equation}
as a spatial point process in $\mathbb R^3$. Each point
$r_{Tk}\in\mathcal M$ is uniformly distributed
within its own``mirror room'' and there is exactly one point per
mirror room. This makes $\mathcal M$ a homogeneous  point process
with intensity 
\begin{equation}
  \label{eq:7}
\varrho_{\mathrm{m}}(r) = 1/V,\quad r\in\mathbb R^3.  
\end{equation}
 Clearly,  $\mathcal M$ is a random point process with  much more
 structure than the familiar spatial Poisson point process. Indeed,  given any
 of the points in the process, all other points are known
 perfectly. In contrast hereto, since the points of a Poisson process are
 independent, knowledge of one point gives no information of the
 presence or location of other points.
 
\begin{figure}
  \centering
\resizebox{1\linewidth}{!}{\includegraphics{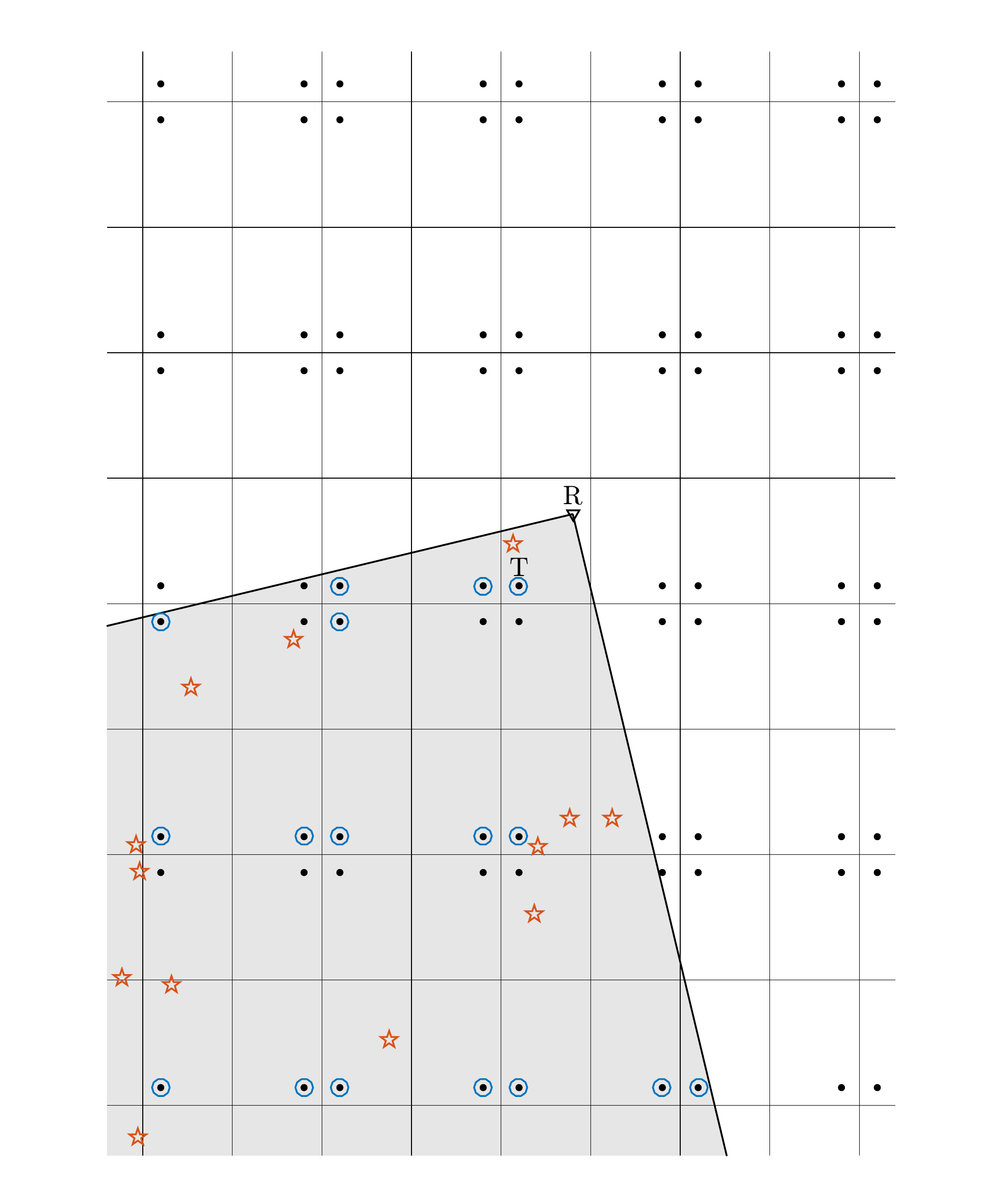}}

  \caption{Realizations of the mirror source process and corresponding Poisson
    approximation. For readability, only sources in mirror rooms with $k_z=0$
    are shown projected to the  $x$--$y$ plane. The gray area indicates the beam coverage of the
    receiver antenna (R). The transmitter antenna (T) is a hemisphere
    oriented  in the direction of the receiver.  The antennas are  at the
    same height.  Black dots: Mirror source positions ($\mathcal
    M$). Blue circles: The mirror sources for which the
    receiver is in the beam coverage ($\mathcal V$). Red stars:
    Poisson approximation ($\mathcal V_{\mathrm{PPP}}$). }
  \label{fig:PoissonApproximation}
\end{figure}

Due to the directive antennas, some of the mirror sources
may not contribute to the received signal, and are hence considered
'invisible'. We consider a path as 'visible' if and only if
both the direction of departure and the direction of arrival reside
within the respective beam supports of the transmitter and
receiver.  Then the set of 'visible' mirror sources reads  
\begin{equation}
\label{eq:9}
\mathcal V = \left\{r \in\mathcal M: 
\frac{r-r_T}{\|r-r_T\|} \in \mathcal
    O_T, 
\frac{r-r_R}{\|r-r_R\|} \in \mathcal
    O_R\right\}.  
\end{equation}
The intensity function of $\mathcal V$ can be derived by noticing  that
due to the assumption of uniformly distributed transmit antenna
orientation, the probability for the antenna of a mirror source to be
oriented toward the receiver, i.e. to have direction of departure within the
beam support of transmitter antenna, is $\omega_T$. Furthermore, a mirror
source only contributes if the direction of arrival is also within the
(deterministic) beam support of the receiver antenna giving the intensity function 
\begin{equation}
  \label{eq:10}
  \varrho_{\mathrm v}(r) = \mathds 1\left(\frac{r-r_R}{\|r-r_R\|} \in \mathcal
    O_R\right)
  \frac{\omega_T}{V}, \quad r\in \mathbb R^3. 
\end{equation}
Fig.\ref{fig:PoissonApproximation} illustrates the two point processes $\mathcal M$
and $\mathcal V$. The process $\mathcal V$ is a subset of
$\mathcal M$ and therefore the points in $\mathcal V$ coincide with
points in $\mathcal M$. 

Relation \eqref{eq:8} maps $\mathcal V$ into  a one-dimensional
point process on the delay axis, i.e.
\begin{equation}
  \label{eq:71}
\mathcal T = \{  \| 
r- r_R\|/c : r\in \mathcal V\}. 
\end{equation}
Using Campbell's theorem, the  mean arrival count for $\mathcal T$ can be derived as 
\begin{align}
\mathbb E[N(\tau)] 
&=\int \varrho_{\mathrm v}( r) \mathds 1 (\| r-r_R\|<c\tau)
 d r\notag\\
&= \frac{4\pi c^3\tau^3}{3V} \cdot \omega_T\omega_R
\cdot \mathds 1(\tau>0).
  \label{eq:72}
\end{align}
As it should, this agrees with \eqref{eq:75} and thus the arrival rate
(intensity function) $\lambda(\tau)$ of
$\mathcal T$ is given in \eqref{eq:63}.

All information needed to evaluate the received signal 
using \eqref{eq:66}, can be collected in form of a marked 
process 
\begin{equation}
  \label{eq:15}
\mathcal {X} = \{(\tau_k,\alpha_k) : \tau_k\in \mathcal T\},
\end{equation}
with the gain given by \eqref{eq:147}.

The three point processes $\mathcal M, \mathcal V$ and $\mathcal T$
and the marked point process $\mathcal X$ all have a structure
reflecting their geometric construction. Given any particular  point in
$\mathcal M$, the whole realization is completely determined. Due to
this  structure, it is very challenging, if at all possible,   to
obtain second or higher order characterizations for
these point processes  \cite{Pedersen2018}. 
 This observation is in line  the well  investigated problem
 in stochastic geometry of counting lattice points
inside a sphere with random center, see e.g. \cite{Kendall1963}. The mean is
known exactly \cite{Kendall1963}, but the asymptotic behaviour for the
deviation from mean, including the variance, is still being investigated; the
standing conjecture in the literature being that the count variable
approaches a Poisson variable as the radius of the sphere
increases \cite{Minami2000,Bourgain2016}. 
This observation, however, motivates our hypothesis that the arrival time
process can be approximated adequately by Poisson point process. To
evaluate the accuracy of such an approximation, we must resort to
simulation studies due to the lack of a higher order characterization.

\subsection{Poisson Approximation for the Mirror Source Process}
\label{sec:turin-model-prop}
To facilitate analysis, we give Poisson approximations for the
processes $\mathcal M$, $\mathcal V$ and
$\mathcal T$. The point process $\mathcal M$ is approximated by a homogeneous
Poisson point process in $\mathbb R^3$  according to 
\begin{equation}
  \label{eq:2}
  \mathcal M \approx \mathcal M_{\mathrm{PPP}} \sim 
  \mathrm{PPP}( \mathbb R^3,\varrho_\mathrm m) 
\end{equation}
where the notation $ \mathrm{PPP}(\cdot,\cdot)$ denotes a Poisson
point process in the specified set and with the specified 
intensity function.  To approximate $\mathcal V$ we account for the
antennas. The transmitter antenna is accounted for by independently thinning 
$\mathcal M_{\mathrm{PPP}}$ keeping points probability $\omega_T$.
The receiver antenna is accounted for by keeping only points with
direction of arrival within the beam coverage.  
This procedure yields a Poisson point process $\mathcal V_{\mathrm
  {PPP}}$ with the same intensity function as $\mathcal V$, i.e.
\begin{equation}
  \mathcal V \approx
  \mathcal V_{\mathrm{PPP}} \sim 
  \mathrm{PPP}( \mathbb R^3,\varrho_\mathrm v).
\end{equation}
Fig.~\ref{fig:PoissonApproximation} gives an example of a realization
of $\mathcal V_{\mathrm{PPP}}$.

Mapping the process $ \mathcal V_{\mathrm{PPP}}$ to the delay axis as in
\eqref{eq:71} gives according to Kingman's mapping theorem
\cite{Kingman1993} a new Poisson point process with intensity
function $\lambda(t)$
\begin{equation}
\label{eq:17}
\mathcal T \approx \mathcal T_{\mathrm{PPP}} = \{  \| 
r- r_R\|/c : r\in \mathcal V_{\mathrm{PPP}}\}\sim \mathrm{PPP}(\mathbb
R, \lambda). 
\end{equation}

Each point in the arrival process $\tau_k\in\mathcal T_{\mathrm{PPP}}$
is marked  independently with a circular symmetric complex gain
$\alpha_k \sim p(\alpha_k|\tau_k)$ giving  a marked Poisson point
process 
\begin{equation}
  \label{eq:37}
  \mathcal X_{\mathrm{PPP}} = \{ (\tau,\alpha) : \tau \in \mathcal
    T_{\mathrm {PPP}}\}. 
\end{equation}
The mark density $p(\alpha|\tau)$ can be chosen in many ways as we
only require that it  is complex circular and has a specified  variance. Owing to
\eqref{eq:67}, the conditional second moment can  be  chosen to ensure
that  the power delay spectrum coincide with the mirror source model as \eqref{eq:12}. 
For example, we may draw independently the complex gain according to a complex circular 
Gaussian pdf with a specified second moment $ \sigma_\alpha^2(\tau) $.
Alternatively, we may draw the magnitude of  $\alpha|\tau$ from  an appropriate fading model 
(Rayleigh, Rice, log-normal, Nakagami-m, etc.) with specified second moment 
and the phase uniformly distributed on $[0,2\pi)$.   The 
specific, however, choice enters in the forthcoming analysis in a way so it is 
straightforward to account for it. We will leave the choice open for 
now to achieve more generally applicable results.

The underlying Poisson process makes the approximation model
analytically tractable and simple to simulate from. The approximations preserve the intensity
functions, i.e. the first order properties of the processes, but no effort was put into to
preserving higher order properties.  From the example in
Fig.~\ref{fig:PoissonApproximation} it  is apparent, that the Poisson
approximation disregards the boundaries of the mirror
rooms and can result in more than one mirror source per room.  This is
a manifest of the fact that a Poisson process does not include
second and higher-order effects, i.e. interaction between points. For
example the process $\mathcal M$ has exactly one point in each mirror
room whereas the approximation $\mathcal M_{\mathrm{PPP}}$ may contain
any non-negative integer number of points in each room. Therefore,
even though the mean counts of the two processes are exactly the same
we expect some approximation error. This error is assessed based on simulations
reported in Section~\ref{sec:simul-stud-exampl}.

\section{Statistical Moments, Cumulants and Kurtosis}
\label{sec:stat-moments-cumul} 

The arrival rate influences the statistical moments of the received
signal. By construction, the proposed approximation model ensures,
that mean is zero and second moments of the received signal matches
the  mirror source model up to a small approximation error. Therefore, to study
differences between these two models, we resort to higher moments.  

The proposed Poisson approximation model permits
derivation of the characteristic and cumulant  generating functionals of the
received signal $y(t)$, as done in Appendix~\ref{sec:appendix:gener-funct}.
From these functionals we can obtain the statistical moments and
cumulants as a function of time.
For the problem at hand it is convenient to first 
compute the cumulants and then
combine these to obtain expressions for necessary moments as described
in \cite{Schreier2010}. 

The cumulants of $y(t)$ are derived in
Appendix~\ref{sec:appendix:gener-funct}.  The odd cumulants
(and moments) vanish due to circularity of $y(t)$. Of particular
interest are the even cumulants of the form
\begin{align}
  \label{eq:96}
\kappa_{n:n}[y(t)]   & = \int |s(t-\tau)|^{2n} P_{2n}(\tau) d \tau,
                       \quad n = 1,2,3\dots 
\end{align}
with the ``$2n$th-order
cumulant-delay spectrum'' defined as
\begin{equation}
  \label{eq:94}
P_{2n}(\tau) =  \mathbb E[  |\alpha|^{2n}| \tau ] \lambda(\tau). 
\end{equation}
  Note that  the second cumulant $P_2(\tau)$ equals
 the power-delay spectrum $P(\tau)$,  defined  in \eqref{eq:67}. 
 Since  $y(t)$ is
 circular, its fourth moment can be obtained as  \cite{Schreier2010}
\begin{align}
  \label{eq:101} 
\mathbb E [|y(t)|^4]& = 
  \kappa_{2:2}[y(t)]+2 \mathbb E [|y(t)|^2]^2.
\end{align}
This relation can be used to compute the  kurtosis-delay
spectrum for $y(t)$ as 
\begin{align}
  \label{eq:29}
  \mathrm{Kurt}[y(t)] &=\frac{\mathbb E [|y(t)|^4]}{\mathbb E [|y(t)|^2]^2}
 =  \frac{\kappa_{2:2}[y(t)] }{\kappa_{1:1}[y(t)] ^2}+2.
\end{align}
The first term on the righthand side is the ``excess kurtosis''  which
is obtained from the kurtosis by subtracting the kurtosis of a
circular complex Gaussian which equals two.

The excess kurtosis depends on the transmitted
signal, the moments of the path gains, and the arrival
rate. In fact,  inspection of \eqref{eq:29} using \eqref{eq:96}  reveals that scaling
the arrival rate by a constant results in an inverse scaling of the
excess kurtosis, that is
\begin{equation}
  \label{eq:23}
  \mathrm{Kurt}[y(t)] -2 \propto \frac{V}{\omega_T\omega_R}.
\end{equation}
The excess kurtosis for small rooms is expected
to be small  and thus close to that of a Gaussian; large
rooms are expected to lead to large excess kurtosis.

Further insight into the relation between arrival rate and the kurtosis delay
profile can be gained  for large bandwidth case.
For a  time-limited transmitted signal with a duration short enough such  that the
product of the $2n$th-order cumulant delay spectrum  is
 nearly constant, we obtain the approximation
\begin{equation}
  \label{eq:105}
  \kappa_{n:n}[y(\tau)] \approx P_{n:n}(\tau)\cdot \int_0^T|s(t)|^{2n} dt,
\end{equation}
and thus
\begin{equation}
  \label{eq:106}
  \mathrm{Kurt}[y(\tau)] \approx
  \frac{1}{\lambda(\tau)} \cdot
  \mathrm{Kurt}[\alpha|\tau] \cdot
\frac{\int_0^T|s(t)|^4dt}{[\int_0^T|s(t)|^2dt]^2} 
 +2
\end{equation}
where $ \mathrm{Kurt}[\alpha|\tau] $ is the kurtosis of $p(\alpha |\tau)$.
In the case, where  the kurtosis of the complex gain is the same for
all delays, the excess kurtosis is approximately proportional to the inverse
of the arrival rate. In the simplifying case that the kurtosis of $\alpha|\tau$ is
independent of $\tau$,  this approximation predicts that  excess kurtosis should
decay quadratically with delay. Thus at larger delays, the excess
kurtosis vanishes, i.e. approaches that of a Gaussian. This is in line
with the intuition provided by the central limit theorem for shot
noise, see e.g \cite{Snyder1991}.  Care should be exercised here---the
intuition is only valid pointwise in $\tau$ and for short signal
pulses.

In simulations or in measurements, the kurtosis delay profile can be
estimated provided a sufficient number of realizations of $y(t)$ are
at hand. The kurtosis can be estimated using standard kurtosis
estimators e.g. by first estimating the fourth and second moments and
inserting in \eqref{eq:29}. Unfortunately, this estimator is biased for
small number of samples. In Appendix~\ref{sec:KurtEstAppendix} we derive an
unbiased estimator for the fourth cumulant of a circular random
variable which we use here to improve the kurtosis estimator.
This allow us to obtain the kurtosis from
simulations even for the mirror source model, where analysis of the
fourth moment is not available. Thus we can compare the simulated
kurtosis of the mirror source model with the results \eqref{eq:23}
and \eqref{eq:106} for the Poisson approximation.

A different application of the kurtosis delay profile is to use for  estimating
parameters of the arrival rate and is thereby a potential tool to
validate the model based on measurement data.    The kurtosis in \eqref{eq:29}, can  be evaluated numerically given a
specific choice of transmitted signal and fitted by a non-linear least squares
approach to the estimated kurtosis in terms of the model
parameters. For this method to be reliable for practical
settings, the expression in \eqref{eq:29} should be modified to
also account for additive noise on the received signal. If the noise
is Gaussian and independent of the signal contribution, this
adjustment amounts to an additive term in $\kappa_{1:1}[y(t)]$ equal to
the noise variance; the fourth cumulant is unaffected by additive
Gaussian noise. Thus, in practice this approach makes it necessary to
estimate the noise variance. With this in mind, the  approximation in
\eqref{eq:106} give an indication of how the sounding signal
should be chosen to estimate the arrival rate accurately: The sounding
signal should have large kurtosis.

 \section{Arrival Times}
\label{sec:arrivalTimes}
In simulation studies, where realizations of arrival times can be
obtained, we may compare the mirror source model and the proposed
approximation model in terms of statistics of their arrival
times, e.g order statistics or interarrival times.




\subsection{Order Statistics}
\label{subsec:prob-distr-arriv}
Order statistics are wellknown and widely used tools within
the field of statistics \cite{Davison2003}. Order statistics can be meaningfully defined for the
inhomogeneous arrival  processes considered here as follows.  
 Since the arrival process $\mathcal T$ is  a one-dimensional point
process it can be arranged in ascending order according to
\begin{equation}
  \label{eq:22}
\tau_{[1]}\leq   \tau_{[2]}\leq   \tau_{[3]}\leq \dots   
\end{equation}
where  the $n$th order statistic $\tau_{[n]}$ is the delay of the
$n$th arrival.  The $n$th order statistic is  unaffected by the observation time interval, i.e. $\tau_{\max}$,
provided it is selected long enough to ensure that (with high
probability) at least $n$ paths arrive within the observation
window. Conversely: if a minimum of $n$ paths have been observed in a
set of measurements, then we should consider order statistics of at maximum order
$n$. Empirical  cdfs for the order statistics are readily obtained in 
simulations by sorting procedure.



The proposed  Poisson approximation permits derivation of the  cdfs of the  order
statistics of arbitrary order.  The derivation begins by observing that the event that the $n$th order
statistic is less than $\tau$ is equal to the event that the region
count $N(\tau)$ is greater than or equal to $n$. Thus,  probability of having more than
$n$ arrivals before delay $\tau$ reads
\begin{align}
\mathbb P(\tau_{[n]}<\tau)  &= 1-\mathbb P(N(\tau) < n)  \notag \\
 &= 1-\sum_{i=0}^{n} \frac{(\tau/a)^{3i}}{i!} \exp(-(\tau/a)^3)
  \label{eq:35}
\end{align} 
with the definition $   a = \sqrt[3]{{4\pi c^3 \omega_T
    \omega_R}/{3V}}$. It is convenient to recast probability in \eqref{eq:35} in terms of  the gamma function $\Gamma(\cdot)$ and the lower incomplete gamma function
$\gamma(\cdot,\cdot)$ as (see e.g. \cite[(10.70)]{Arfken2001}) 
\begin{align}
\label{eq:5}
\mathbb P(\tau_{[n]}<\tau)= \frac{\gamma(n,(\tau/a)^3)}{\Gamma(n)} = F(\tau;a,3n,3).
\end{align}
Here, $F(\cdot; a,3n,3)$ is a generalized gamma cdf
\cite{Stacy1962} for which the  moment generating function reads
\begin{equation}
  \label{eq:19}
  M_{[n]}(\nu) = \sum_{r=0}^{\infty}\frac{(\nu a)^r}{r!} \cdot \frac{\Gamma(n+r/3)}{\Gamma(n)}.
\end{equation}
From the generating function it is straight-forward to identify the
$r$th moment of the $n$th order statistic.

\subsection{Interarrival Times}
\label{sec:interarrival-times}
The distribution of time intervals between the arrival of
multipath components, called the interarrival times, has been used 
by several authors as a means to fit and validate stochastic channel
models, e.g. \cite{Spencer2000,Chong2005,Gustafson2014}.  If  the
arrival times form a homogeneous Poisson process, the
interarrival times are exponentially distributed. Unfortunately, as
we show in the following,  this distribution of interarrival times is
not well defined in the  inhomogeneous case considered  here.

We first derive the distribution of interarrival time, given that
point arrive at delay $\tau_k$. The probability that no path arrives
between delay $\tau_k$ and 
$\tau_k+\delta$ for fixed interarrival time $\delta>0$ follows from the
probability that no points from  $\mathcal T$ fall in the  interval
$(\tau_k,\tau_k+\delta)$. This is the same as the probability for the
interarrival time, given $\tau_k$ is greater than $\delta$.  Since $\mathcal T$ is a Poisson point
process,  the probability for the interarrival time less than $\delta$ reads
\begin{align}
\mathbb       P(\tau_{k'} -\tau_{k}   <\delta|\tau_{k}) =&
1-\mathbb       P(N(\tau_{k}+\delta) -N(\tau_{k})   = 0|\tau_{k}) \notag\\
  =& 1 - \mathds
                      \exp(-\Lambda(\tau_k+\delta
                                     )+\Lambda(\tau_k)) \notag\\
 =& 1- \exp(-\frac{(\tau_k+\delta)^3-\tau_k^3}{a^3}) , \label{eq:104}
\end{align}
$\delta > 0$. This is a well defined cdf: with $\tau_k = 0$, this is a Weibull cdf with with shape
parameter three and scale parameter $a$; for $\tau_k>0$ it is a
shifted and  truncated Weibull cdf with the same parameters. 

To obtain the (unconditional) distribution of interarrival times the
delay $\tau_k$ should be averaged out. 
By definition of the Poisson process, the arrival times falling in
this interval are iid.\ with pdf 
\begin{align}
  p(\tau_k) &= \frac{\lambda(\tau_k)}{\int_0^{\tau_{\max}} \lambda(\tau) 
    d\tau}\cdot \mathds 1(0\geq\tau_k\geq \tau_{\max})\notag \\
 &=\frac{3\tau_k^2}{\tau_{\max}^3}\cdot \mathds 1(0\geq\tau_k\geq
   \tau_{\max})
  \label{eq:30}
\end{align}
This pdf is well defined for finite $\tau_{\max}$,  but for
$\tau_{\max}\rightarrow \infty$ the denominator in \eqref{eq:30} diverges and
the pdf is zero for finite $\tau_k$.   The expectation of \eqref{eq:104} with respect to \eqref{eq:30} can now be carried out,
e.g. using symbolic computation software. The
result, which we omit here due to its length,  depends on the value of 
$\tau_{\max}$. In particular, for  infinite observation interval,  the
resulting distribution of interarrival times is degenerate with all
probability mass at zero. 

We remark that the notion of interarrival times appears to be  much more
involved for inhomogeneous arrival processes potentially leading to 
misinterpretation. As an example, in
\cite{Chong2005} it was observed that interarrival times were not 
exponentially  distributed as it should be for a homogeneous Poisson
process. The authors concluded that a homogeneous Poisson process
was inadequate to model the data and proposed instead to model
the interarrival times as a mixture of two exponential distributions which fits the measurement
data well. Although unnoticed in the \cite{Chong2005}, or the  
comments raised in \cite{Nadarajah2008},  this modification replaces  
the homogeneous Poisson process by a renewal process, see e.g. \cite{Davison2003}. A renewal  
processes is  specified by the distribution of the interarrival times  
and have constant arrival rate by construction.  The renewal process used in \cite{Chong2005} has in a total of  
three parameters and it is therefore unsurprising that it fits the  
data much better than the constant rate model with only one
parameter.  Moreover, since the interarrival times of an inhomogeneous Poisson process  does not
 have to be exponentially distributed, the observation from
\cite{Chong2005} of non-exponential interarrival times does not
contradict that the arrival process is a Poisson process. The only
safe conclusion is that the arrival  process is not homogeneous Poisson.



\subsection{Residual Power after Removing Dominant Paths}
In measurements, the received signal $y(t)$ can be obtained, but the
points in the arrival process are `hidden' and must be first extracted.
The problem of extracting delays and amplitudes for multipath
models has received a tremendous amount of research attention, and
many good estimation techniques exist, e.g.
\cite{Hogbom1974,Fleury1999}. These techniques tend  to work well
for clearly separated multipaths and when the total number of multipaths
are low and known. Nevertheless, this is likely not the case for the
inroom scenarios considered here. It is also possible to account for
``diffuse components'' in the estimator \cite{Thoma2004}, but to do
so, we should be able to distinguish between specular and diffuse
components. In the light of the gradual specular-diffuse transition
predicted by the mirror source model, such a split seems unnatural. To
apply the estimators \cite{Hogbom1974,Fleury1999,Thoma2004} one
needs to set a number of multipath components to extract. This setting
is critical since the power of the residual (the unresolved part of
the received signal) depends on it. In the following, we use the order
statistics to predict the residual power as a function of this setting.

We consider the ideal case where the
multipath components are extracted one by one according to their
ordering in delay. This enable us to use the derived results for the
order statistics as follows. Denote by $P_{[n]}$ the mean power
contributed by paths with delay greater than $\tau_{[n]}$.  Then the set of arrival
delays exceeding $\tau_{[n]}$ can be written as 
\begin{equation}
  \label{eq:16}
\mathcal T_n = \mathcal T_{\mathrm {PPP}}\cap (\tau_{[n]},\infty) = \{\tau_{[n+1]}, \tau_{[n+2]},\dots \} .
\end{equation}
The total power of paths with delay greater than $\tau_{[n]}$ reads
\begin{align}
  \label{eq:14}
P_{[n]} &= 
\mathbb E\bigg[ \int  \bigg|\sum_{i: \tau_i \in \mathcal T_n} \alpha_i s(\tau-\tau_i)\bigg|^2
  d\tau\bigg| \tau_{[n]}\bigg]\\
& = E_s  \mathbb E[ \sum_{\tau\in \mathcal T_n}
         \sigma^2_\alpha(\tau) | \tau_{[n]}] 
\end{align}
where $E_s = \int |s(\tau)|^2d\tau $ and zero
mean uncorrelated path gains are assumed. Since $\mathcal
T_{\mathrm{PPP}}$ is a Poisson point process, its points are
independent. Therefore, the set $\mathcal
T_n|\tau_{[n]}$ is a Poisson point process  with intensity function
$\lambda(\tau) \mathds 1(\tau>\tau_{[n]})$. Then by invoking
Campbell's theorem, \eqref{eq:12} and~\eqref{eq:67} we obtain
\begin{align}
P_{[n]} & = E_s\int_{\tau_{[n]}}^\infty P(\tau) d\tau \\
& = E_s \frac{4\pi c T}{3V} \mathbb \exp(-\tau_{[n]}/T).
\end{align}
Taking the expectation with respect to $\tau_{[n]}$ yields
\begin{align}
  \label{eq:20}
\mathbb E[ P_{[n]} ] & = E_s \frac{4\pi c T}{3V} M_{\tau_{[n]}}(-1/T).
\end{align}
The relative residual power, i.e. the ratio of the residual power and the total power reads 
\begin{align}
\label{eq:34}
\mathbb E[ P_{[n]} ] / P_{tot} & = M_{\tau_{[n]}}(-1/T)   
\end{align}
This ratio is unity for $n=1$, but vanishes  for large $n$ at a decay
rate determined by the ratio $a/T$. Using
\cite[(5.11.12)]{Olver2010}, we see that the residual power has an
asymptote given as
\begin{equation}
  \label{eq:21}
  M_{[n]}(-1/T) = \exp(-\sqrt[3]{n}a/T), \qquad n \rightarrow \infty. 
\end{equation}
The residual power  decays more slowly in
terms of $n$ for smaller rooms than  for larger rooms. Furthermore, the decay is
affected by the antennas. For more directive antennas, the decay is
faster because the power is concentrated on fewer multipath
components.




\section{Simulation Study}
\label{sec:simul-stud-exampl}
\begin{table}
  \centering
  \caption{Simulation Settings }
  \begin{tabular}{lc}
\toprule
  Room dim., $L_x\times L_y\times L_z $& $5\times 5\times 3\, \cubic\meter$\\
  Reflection gain, $g$ & $0.6$\\
 Center Frequency & $60\,\giga\hertz$\\
 Bandwidth, $B$& $2\,\giga\hertz$\\
 Wavelength, $l_c$ & $30\, \milli\meter$\\ 
 Speed of light, $c$ &$3\cdot 10^8\,\meter\per\second$\\ 
Maximum delay, $\tau_{\text{max}}$ & $100\,\nano\second$\\
Rate in constant rate model, $\rho_0$ & $ \omega_T\omega_R  \cdot 150 / \tau_{\text{max}}$\\
Transmitted signal, $s(t)$ & Hamming pulse\\
Antennas & Sph. Cap Sector\\
No. Monte Carlo Runs & $10^4$\\
\bottomrule 
\end{tabular}
  \label{tab:simulationSettings}
\end{table}

 The accuracy of the proposed Poisson approximation model is 
evaluated by means of Monte Carlo simulations of the following three models: 
  \begin{enumerate}
\item  \emph{MS:} The mirror source model defined in Section~\ref {sec:mirrorSourceTheory}
  with uniformly random antenna positions and orientations.
\item \emph{Proposed:} The inhomogeneous Poisson approximation defined in
  Section~\ref{sec:turin-model-prop}. The method used to generate the inhomogeneous 
Poisson process is described in Appendix~\ref{sec:SimAppendix}. The gains
are assumed complex Gaussian.
\item \emph{Constant Rate:} A homogeneous Poisson model with constant
  arrival rate $\rho_0$, but with the same delay power spectrum as in
  Subsection~\ref{sec:turin-model-prop}. The gains are assumed complex Gaussian.
  \end{enumerate}
The constant rate model is included to contrast the proposed inhomogeneous 
model and the homogeneous case assumed in
e.g. \cite{saleh,Haneda2015}.  It should be noticed, however, that the
effect of the antennas has not hitherto been included in the
constant arrival rate models. We do so here to illustrate how the
antenna effect would enter in the constant rate model. Simulation results
of the models are compared in terms of the statistics investigated in
Sections~\ref{sec:stat-moments-cumul} and~\ref{sec:arrivalTimes}. In
addition we simulate mean delay and rms delay spread to illustrate how
the inhomogeneous model impacts these parameters considered in design and
evaluation of communication systems.

The simulation setup is the same as in \cite{Pedersen2018} with
settings as specified in Table~\ref{tab:simulationSettings}.  The
transmitted signal $s(t)$ is a Hamming pulse with the considered frequency
bandwidth. To achieve finite computational complexity, we simulate only
components with a delay up to a maximum delay denoted by
$\tau_{\text{max}}$.   To illustrate the impact of the beam coverage of the antennas, we
 consider identical lossless spherical cap sector antennas as defined in
 \cite{Pedersen2018}. For this type of antenna, $\omega$ specifies the
 response: $\omega=1$ yields the isotropic antenna response and
 $\omega=0.5$ yields a hemisphere antenna.



\subsection{Example Realizations}

\begin{figure*}
\centering\scriptsize
\begin{tabular}[c]{cc}
 \resizebox{0.45\linewidth}{!}{\includegraphics{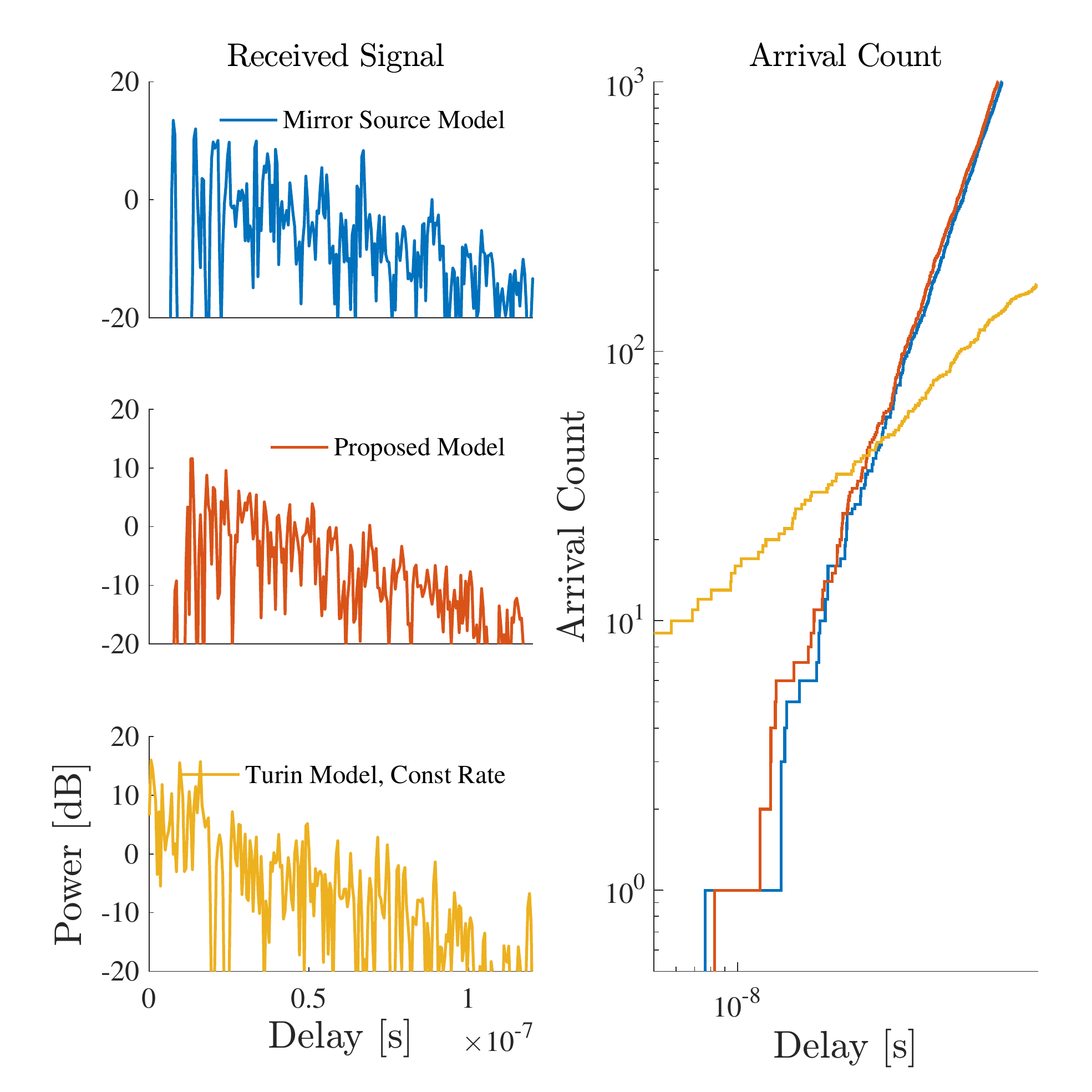}} &
\resizebox{0.45\linewidth}{!}{\includegraphics{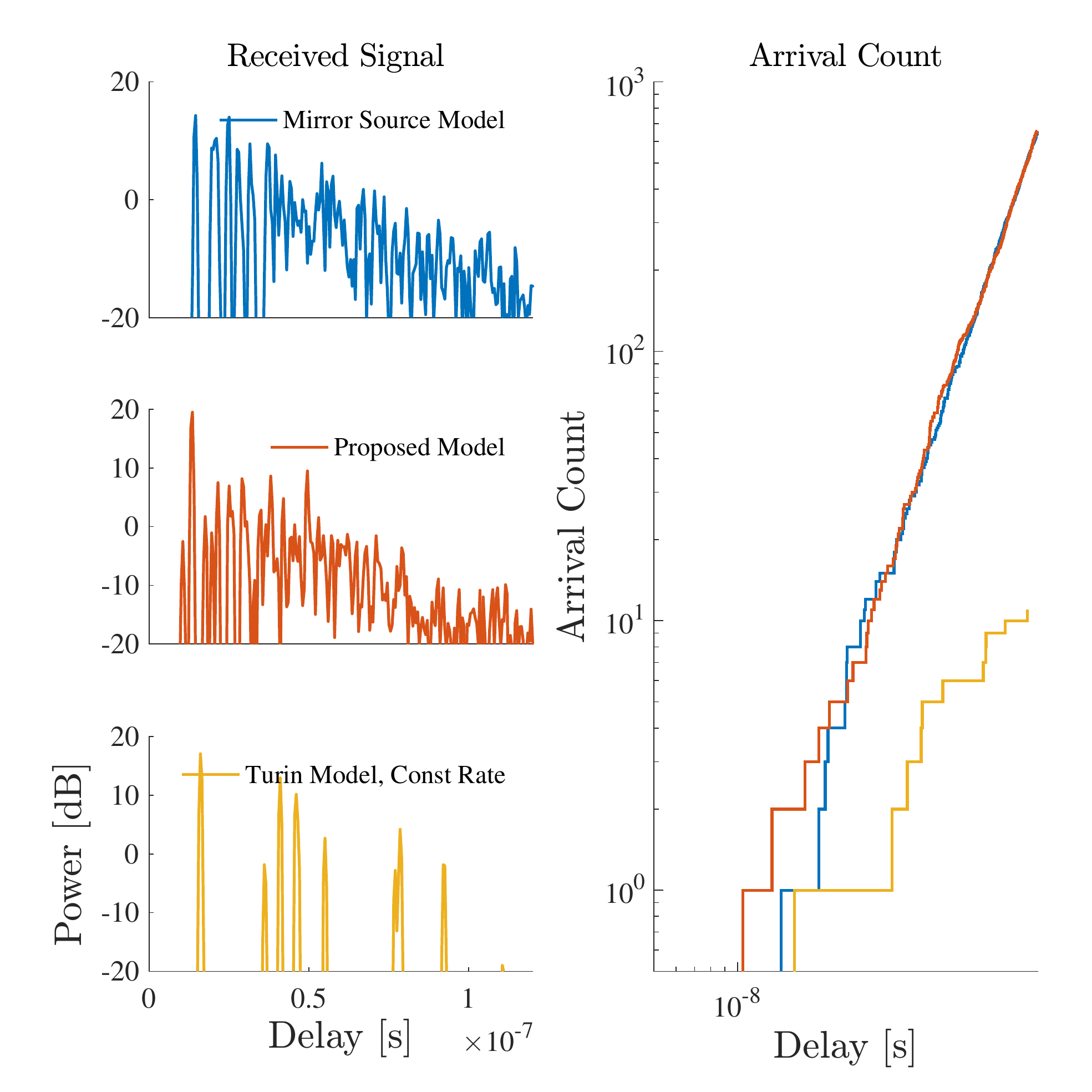}}
  \\
(a) Isotropic Antennas $\omega_T=\omega_R = 1$  &
 (b) Hemisphere Antennas $\omega_T=\omega_R = 0.5$  
\end{tabular}
\caption{Example realizations of the received signal (magnitude
    square)  and corresponding arrival counts for (a) isotropic  and
    (b) hemisphere  antennas. }
\label{fig:realizations}
\end{figure*}

Fig.~\ref{fig:realizations} gives examples of individual realizations 
for the squared magnitude of the received signal. The received signals for the mirror source model and 
the proposed approximation model both exhibit a specular to diffuse 
transition, i.e. early well separated specular components are
succeeded by a gradually denser diffuse tail. This effect is not replicated by the  constant rate 
model which is either ``constantly sparse'' or ``constantly
dense''.

Fig.~\ref{fig:realizations} also show the arrival counts for the three
models. The arrival counts are not observable in a measurement, but
are easy to obtain in simulations.  As expected,  the arrival counts for the three models
fluctuate about their respective theoretical mean count. The mirror
source model and the proposed approximation produce similar
realizations of arrival count versus  delay while the constant rate
model differ.  Moreover, as predicted, the count for the isotropic
antennas is four times higher than that obtained with the hemisphere
antennas.

\subsection{Power and Kurtosis of the Received Signal}
\begin{figure*}
  \centering
\centering\scriptsize
\begin{tabular}[c]{cc}
\resizebox{0.45\linewidth}{!}{\includegraphics{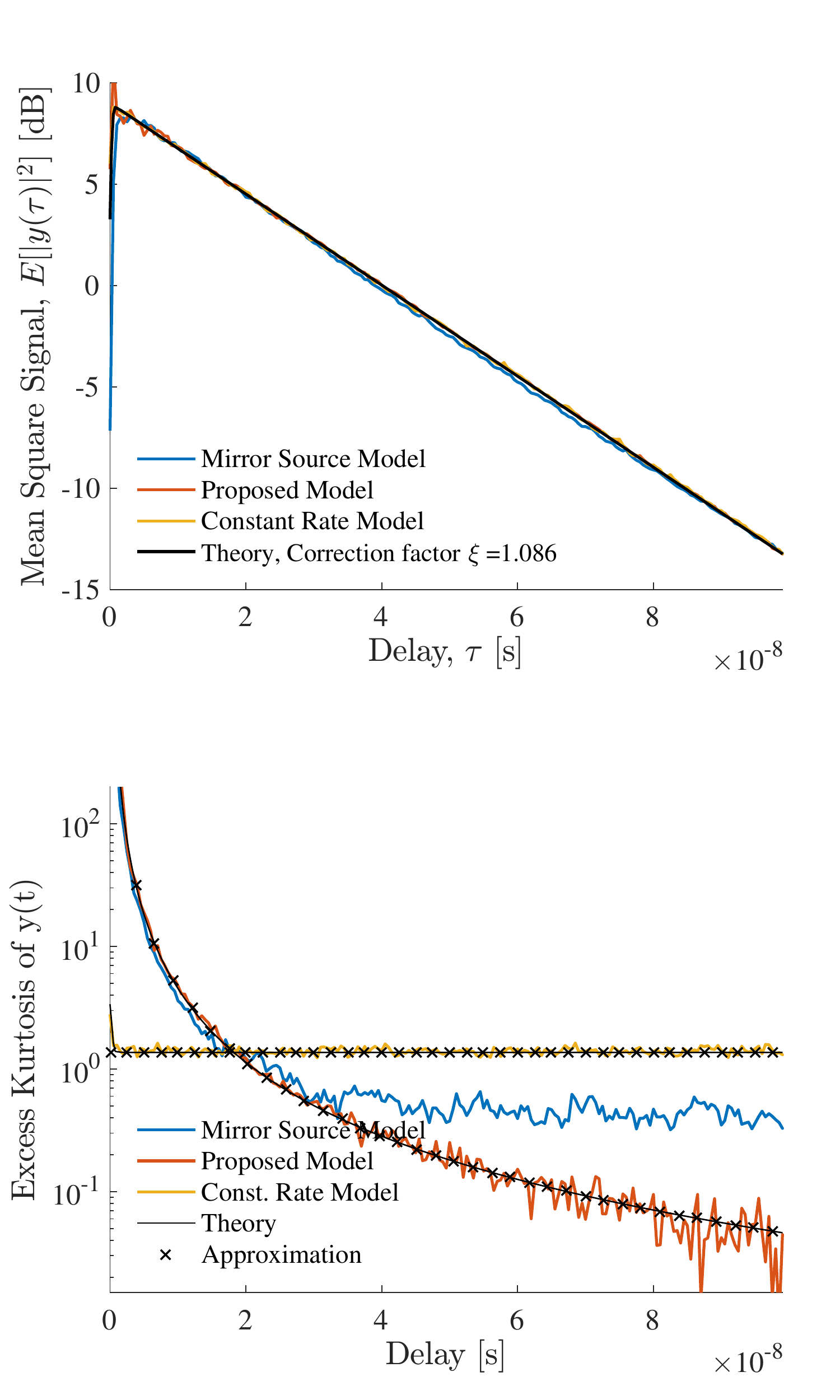}}& 
\resizebox{0.45\linewidth}{!}{\includegraphics{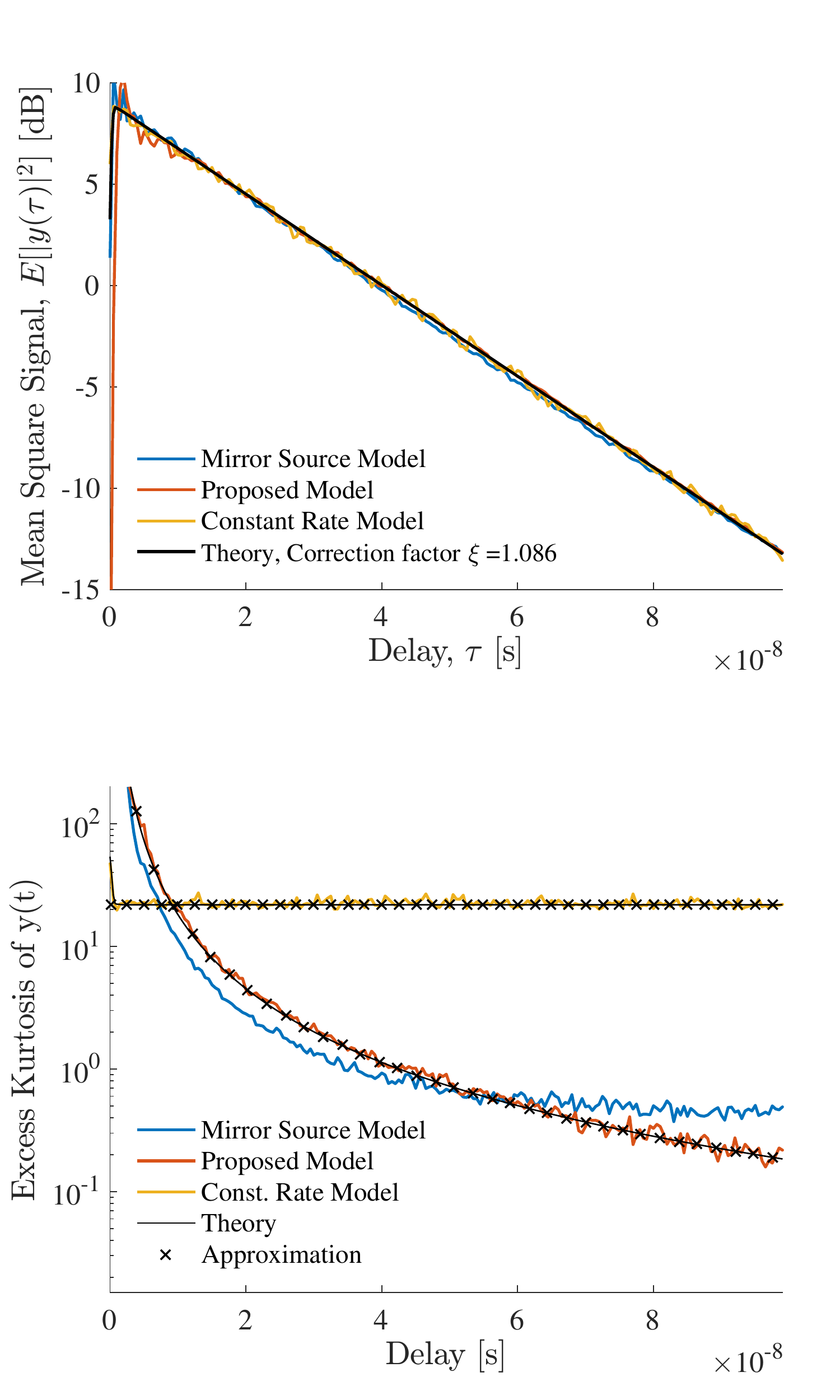}}
  \\
(a) Isotropic Antennas $\omega_T=\omega_R = 1$  &
 (b) Hemisphere Antennas $\omega_T=\omega_R = 0.5$  
\end{tabular}
  \caption{Power and kurtosis of the received signal obtained from  simulation,
    theory and approximation for (a) isotropic  and
   (b) hemisphere  antennas.}
  \label{fig:pds}
\end{figure*}

The upper panels of Fig.~\ref{fig:pds} shows the simulated
expected received power versus delay for the three models along
with the theoretical value computed using
\eqref{eq:93}.  The theoretical and simulated curves agree well for
all three models with only minor deviations in the early parts of the
spectra due to the applied Monte Carlo simulation technique.  Thus,
the simulations confirm that all three models have the same power
delay spectrum and clearly exemplifies that models with very different
arrival rates can indeed have identical second order statistics. 
Comparing the results for isotropic and hemisphere antennas it appears
that the antenna directivity does not affect the power delay spectrum. In addition, we see that for
the considered simulation setup,  the high-bandwidth
approximation obtained using \eqref{eq:105} with $n=1$ is very
accurate with some minor discrepancies at small delays.

Excess kurtosis delay spectra are reported in the lower panels of
Fig.~\ref{fig:pds}. The theoretical curves computed using
\eqref{eq:29} are close to to the large  bandwidth approximations
obtained  using  \eqref{eq:105} and~\eqref{eq:106}.
The theory predicts the kurtosis to increase by a factor of four by
replacing the isotropic antennas by hemisphere antennas. It appears that this shift is correctly represented in all three models.  The simulations for the mirror source model agrees well with simulations of the  proposed
model. In particular this is true for the early part of the response
which carries the most signal power.   At later delays, however, the
mirror source model deviates somewhat from the  proposed. 
The deviation is  caused in part by the small discrepancy in the model
of the second moment, and in part due to the fact that the gain variables of the mirror source
 model is approximated by a Gaussian random variable. The discrepancy
 is furthermore accentuated by the logarithmic of the second axis.
The curve for constant rate model clearly differs from the two others---as expected its
kurtosis delay profile is constant. From this simulation it is
apparent that  models with identical delay power spectra may differ
significantly in their predictions of the kurtosis delay spectrum.

\subsection{Order Statistics of Arrival Times}

\begin{figure*}
  \centering
\centering\scriptsize
\begin{tabular}[c]{cc}
 \resizebox{0.5\linewidth}{!}{\includegraphics{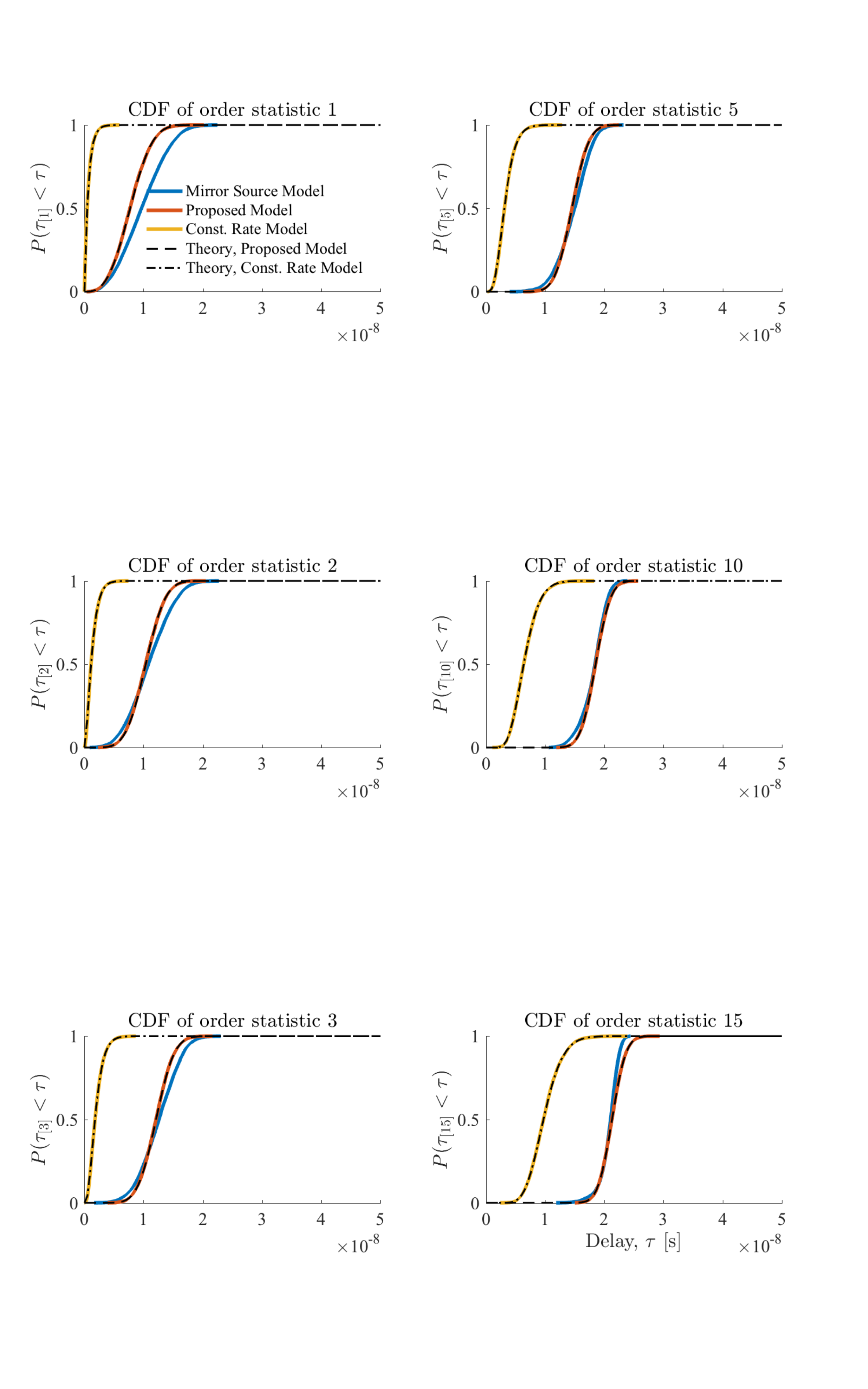}}&
 \resizebox{0.5\linewidth}{!}{\includegraphics{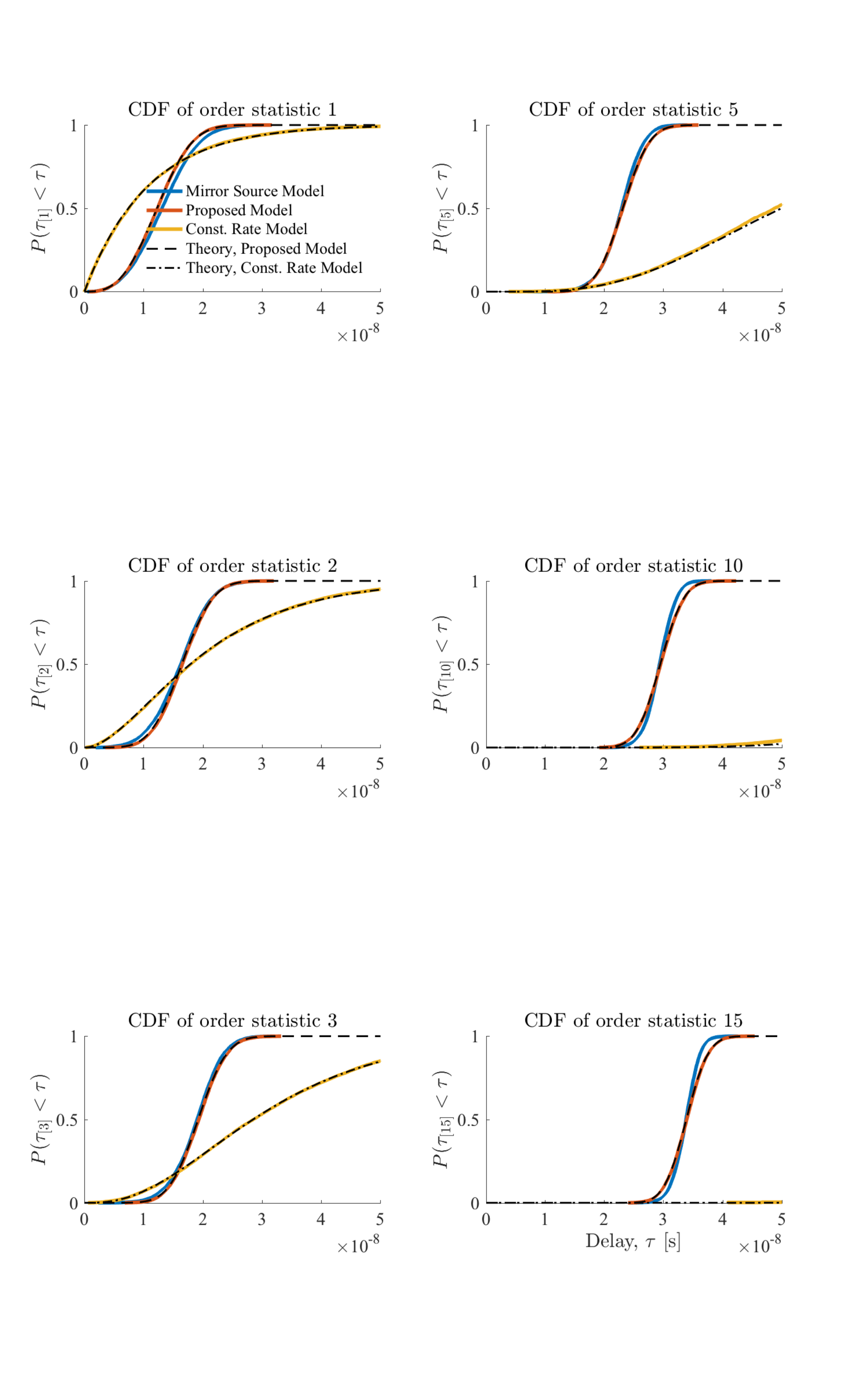}}
  \\
(a) Isotropic Antennas $\omega_T=\omega_R = 1$  &
 (b) Hemisphere Antennas $\omega_T=\omega_R = 0.5$  
\end{tabular}
  \caption{Empirical cumulative probability for the order statistics
    of the three models with (a) isotropic and (b) hemisphere
    antennas. The theoretical cdfs given in
    dashed lines for the proposed and the constant rate models fall on
    top of the simulated curves.}
  \label{fig:orderStatistics}
\end{figure*}



The simulated order statistics for the arrival times reported in
Fig.~\ref{fig:orderStatistics} give rise to a number of observations.
As to be expected, for all models the cdfs of the order statistics move to the right as
the order increases. Moreover, the slope of the cdf (related
to the variance of the pdf) is steeper for the  isotropic
antennas than for the hemisphere antennas. This indicates that 
more directive antennas lead to a larger spread of the order
statistics. For all considered order statistics, the proposed model
captures more accurately the shape of the cdf than the constant rate
model.  This indicates that to accurately model the order statistics of the 
arrival process, it is important to model the arrival rate
properly. In the considered case, the constant rate model is not
appropriate. 
The deviations between the proposed model and the mirror source model
are relatively minor and most significant in the first few  order statistics and in the case with
isotropic antennas. This effect shows that the approximation of
the mirror source process in terms of Poisson process is most
significant for isotropic antennas at the early delays, including the
line-of-sight component. 

\subsection{Residual Power After Removing Paths}

\begin{figure}\scriptsize
   \centering   \resizebox{0.8\linewidth}{!}{\includegraphics[trim={0 10cm 0 0},clip]{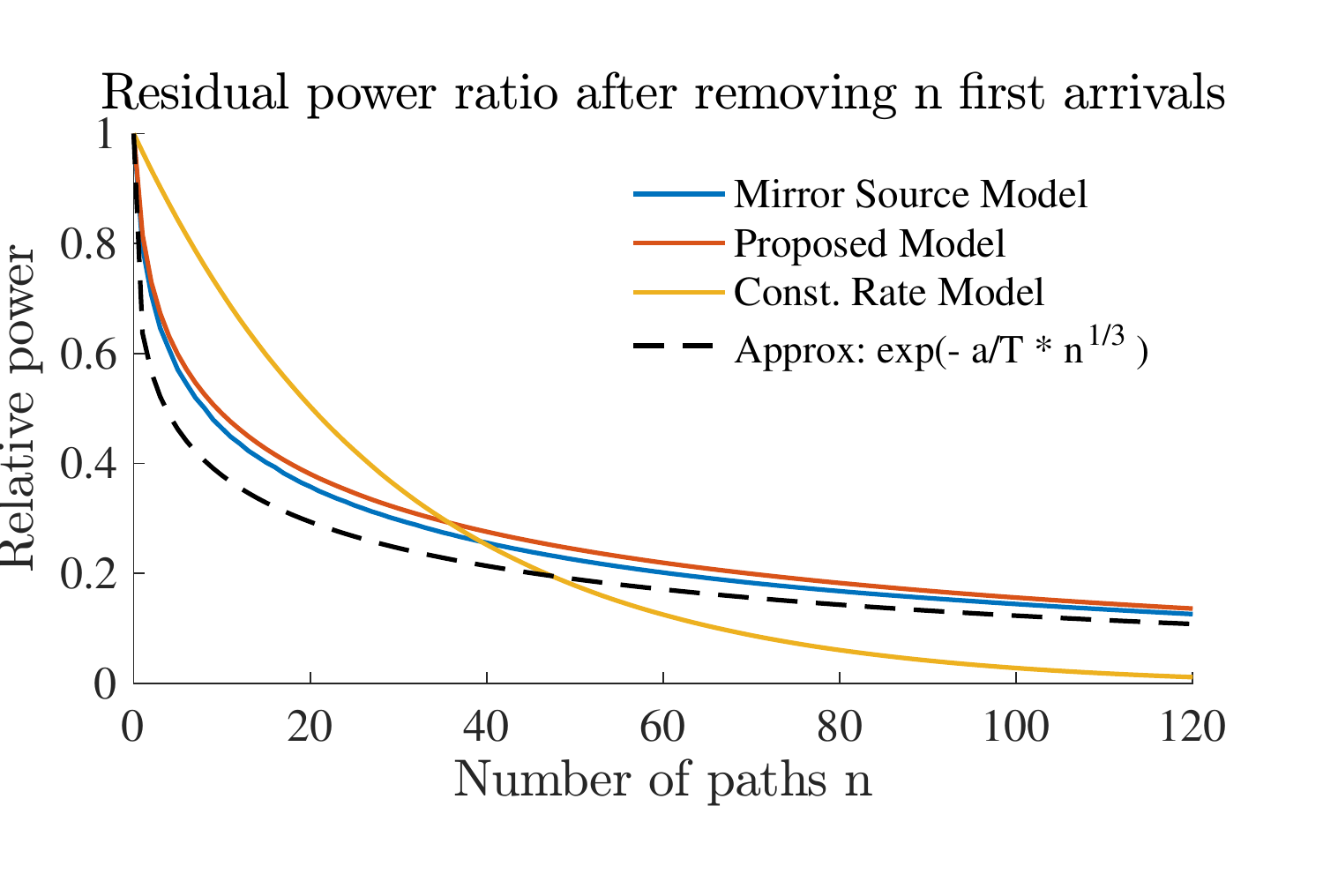}}\\
(a) Isotropic Antennas $\omega_T=\omega_R = 1$  \\[1ex]
   \centering   \resizebox{0.8\linewidth}{!}{\includegraphics[trim={0
       10cm 0 0},clip]{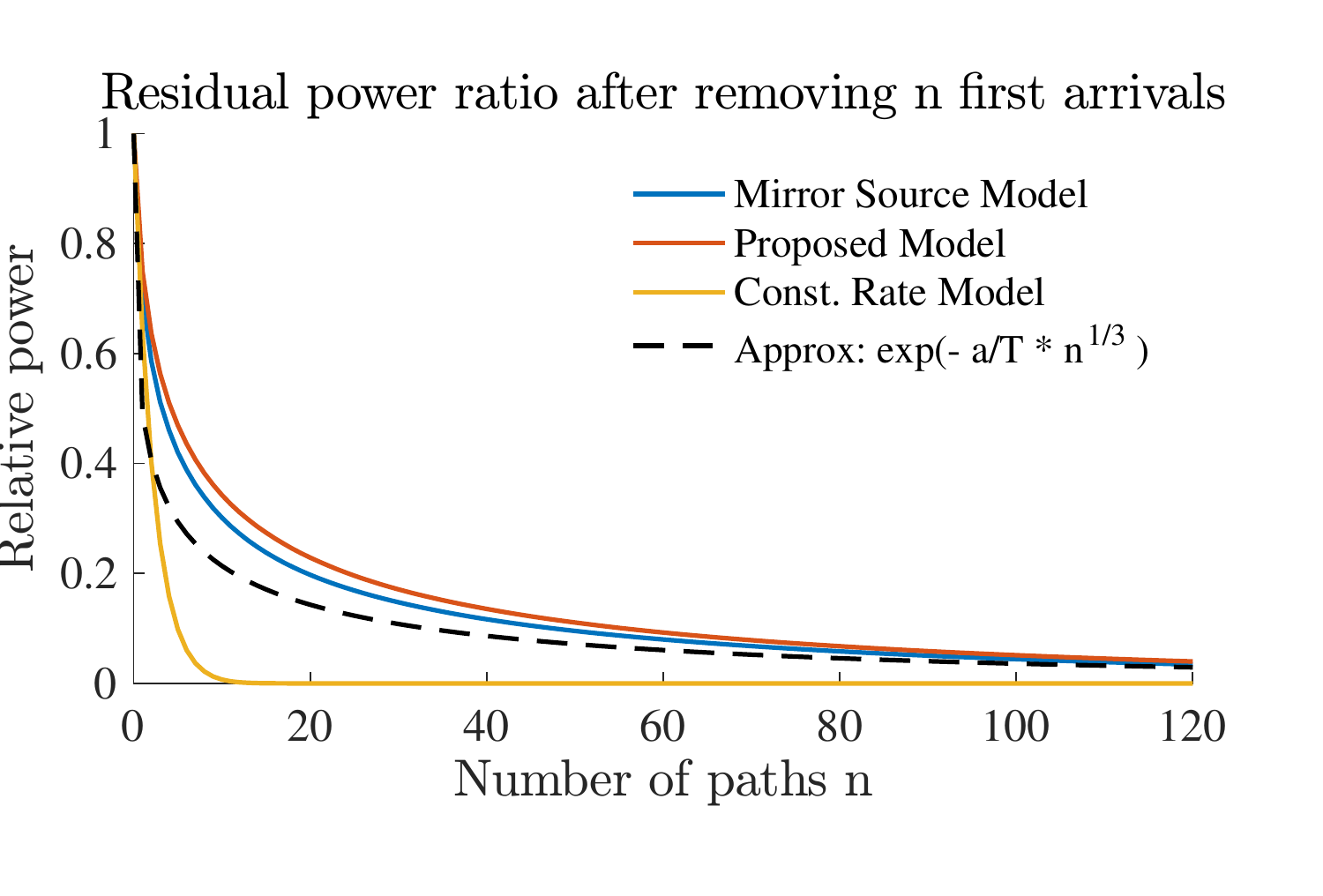}}\\[1ex]
 (b) Hemisphere Antennas $\omega_T=\omega_R = 0.5$  
   \caption{The residual power relative to the total power for the
     three models after removing $n$ first paths for (a) isotropic
    and (b) hemisphere antennas. The approximation \eqref{eq:21} is given in dashed line. }
   \label{fig:residualPower}
\end{figure} 

Fig.~\ref{fig:residualPower} shows the relative residual power after removing
$n$ first arrivals.  The residual power obtained with the  proposed model fits well the results obtained  from the
mirror source model. This is to be expected in the light if the close match in
order statistics observed in Fig.~\ref{fig:orderStatistics}.
Also, the approximation computed in \eqref{eq:21} 
predicts well the trend of the received power. The constant rate
model  differ clearly from the two other models.

By comparing the upper and lower panels of
Fig.~\ref{fig:residualPower} it is clear that the antenna
characteristics affect the residual power for all three models.   In the proposed model, the residual power depends
only on the ratio $a/T$, it varies with room size, antenna characteristics,
and the reverberation time.  This observation is relevant in particular
in connection with approximating the received signal using only a
fixed number of multipath components such as commonly done using  a high-resolution
multipath estimators  \cite{Hogbom1974,Fleury1999, Thoma2004}.
This model prediction is in agreement with the recently published measurement results \cite{Saito2017} where the residual
power after removing specular components is observed to decay at different rates
for differently sized rooms.

\subsection{Instantaneous Mean Delay and RMS Delay Spread}
\begin{figure*}
  \centering
\centering\scriptsize
\begin{tabular}[c]{ccc}
 \resizebox{0.4\linewidth}{!}{\includegraphics{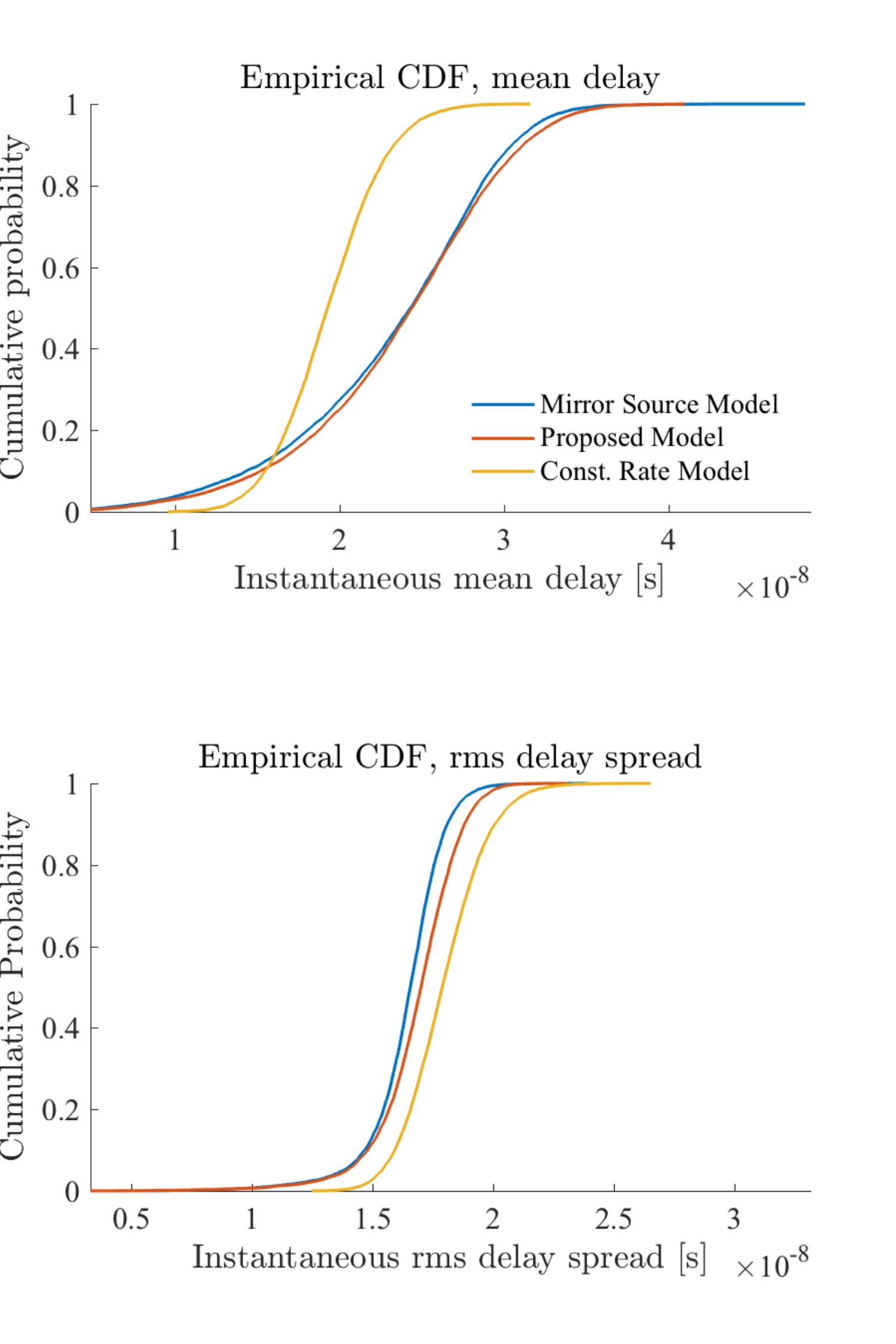}}&&
  \resizebox{0.4\linewidth}{!}{\includegraphics{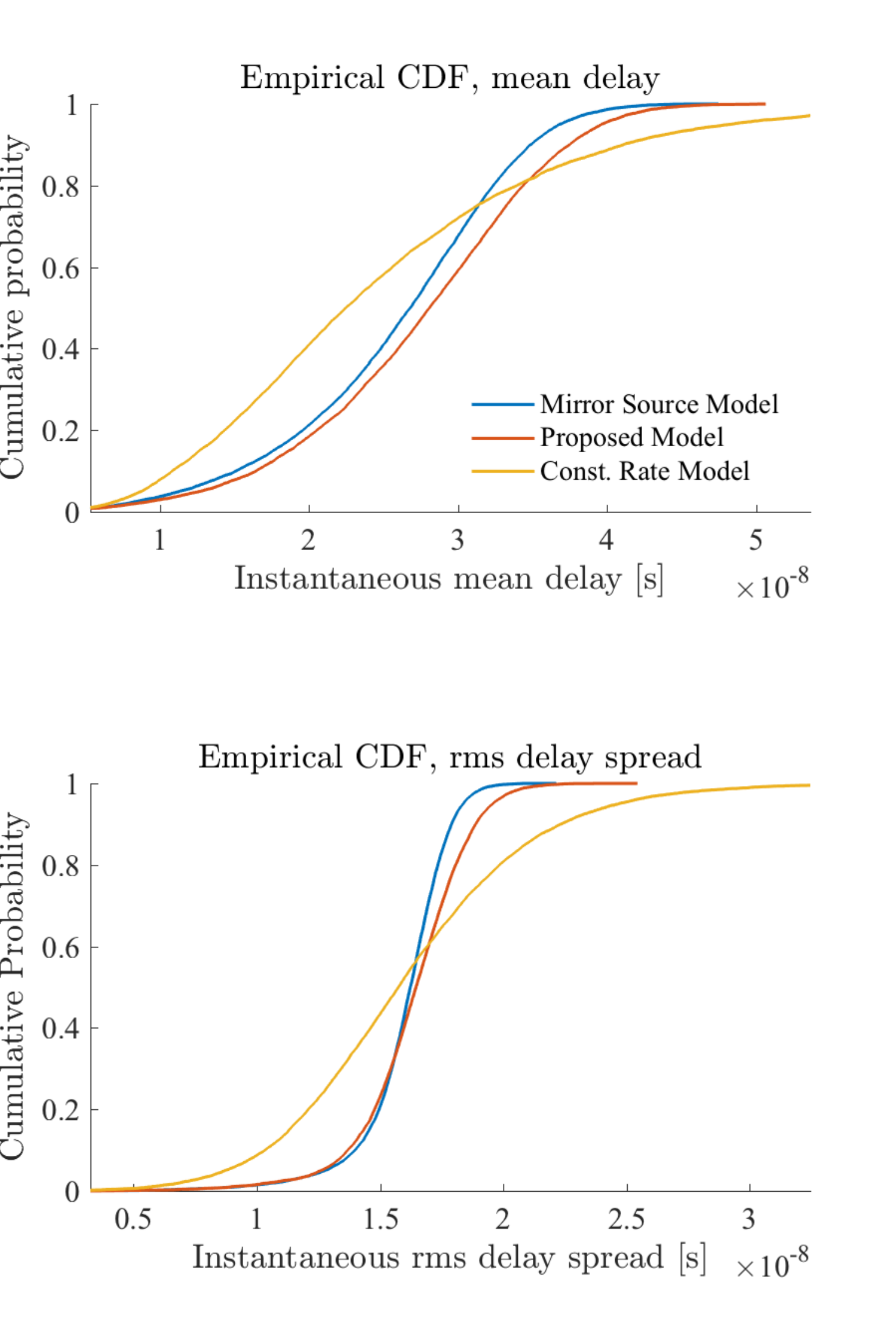}}
\\
(a) Isotropic Antennas $\omega_T=\omega_R = 1$  &&
 (b) Hemisphere Antennas $\omega_T=\omega_R = 0.5$  
\end{tabular}
  \caption{Empirical cumulative probability for instantaneous mean
    delay and  rms delay spread for (a) isotropic and (b)
    hemisphere antennas. }
  \label{fig:rmsDelay}
\end{figure*}

The distributions of instantaneous mean delay and
rms delay spread are important for design of radio communication
systems since numerous aspects of
the system performance is characterized via these
parameters. Therefore, we compare the considered models by comparing
the resulting distribution of these parameters.  Theoretical analysis
of mean delay and rms delay spread 
is beyond the scope of this contribution, so we only report simulation
results.   Here, the mean and rms  are
computed as respectively the first and centered second temporal
moments of each realizations of $|y(t)|^2$ (thus including 
the effect of the transmitted pulse).   

The empirical cdfs for mean
delay and rms delay spread reported in Fig.~\ref{fig:rmsDelay}.
It appears that proposed model is able to mimic the effects of the mirror source
model well enough to accurately capture the distributions mean delay and rms delay
spread.  This is not the case for the constant rate model. All three models predict a shift of the
curves as the directivity of the antennas change.

It should be   noted that  in the considered case, the three models have identical  
power delay spectra, and thus the differences between the proposed
inhomogeneous and the constant models  only stem from  
differences in higher moments which are controlled by their arrival rate. 
We conclude from these observations that accurate modeling of the arrival rate
is a necessity to correctly model the  distribution of instantaneous
mean delay and rms delay  spread.

\section{Discussion}
\label{sec:discussion}
The proposed stochastic model is based on the mirror source analysis
presented in \cite{Pedersen2018}. Thus, the proposed model originates
from an approximation of the mirror source theory for a very
simplistic scenario where transmitter and receiver are placed in a
rectangular room with perfectly flat walls void of other objects.
Certainly, in most realistic scenarios, the walls will be imperfect
due to doors, windows, heating devices, ventilation ducts, light
fixtures and other such details. In addition, other objects in the
room add to the complexity of the propagation environment. Therefore,
the mirror source model should in itself be considered as an
approximation of any real propagation environment.

We do expect, however, that since the very major elements of the
inroom scenario, namely the walls, floor and ceiling are accounted
for, the model can be used to qualitatively predict some effects that
might occur in more realistic cases. We conjecture that if the
scenario is made more complex, e.g by considering a furnished room,
a number of mirror sources should be added which leads to an even
faster growth of the arrival rate. This will speed up the diffusion
process and result in an even faster decay of kurtosis delay profile.

The present contribution has focused on the theoretical analysis of
the proposed model rather than its experimental validation. As
discussed in \cite{Pedersen2018}, the power delay spectrum agrees with
a model which has been previously been experimentally validated.
Other predictions of the proposed approximation model have, however,
not yet been compared to measurement data.  We comment on
validation of the model  in the following.

The predictions related to the moments of the received signal can be
validated as they are easy to relate to measurement data. In
particular, the kurtosis delay profile could be estimated using using
the estimator in Appendix~\ref{sec:KurtEstAppendix}. While this seems
straight-forward, we should bear in mind that reliable estimation of
higher moments (here the fourth moment) usually calls for a large
number of measurements. The presence of noise in the measurements may
impair the estimation accuracy of especially the late part of the
kurtosis delay profile.  Therefore, we suggest that the robustness and
noise-sensitivity of the estimator of the kurtosis delay profile
should be investigated in more detail prior to applying these
estimators.

To validate the model based on arrival delays and complex gains, (a
part of) the marked point process $\mathcal X$ should be estimated.
To this end, it is necessary to apply high-resolution estimators such
as \cite{Hogbom1974,Fleury1999,Thoma2004}.  Such estimators, however,
detect more easily multipath components with short delays which tend
to have the strongest power and be better separated.  Hence more
signal components are missed by the estimator in the parts of the
response where the density is the largest and the gains are the
weakest.  This effect can be considered as censoring of the
observation \cite{Karttunen2017} which may severely bias 
statistics based on estimates of arrival times.

As mentioned in Section~\ref{sec:interarrival-times}, it is a the
widespread practice of calibrating and validating models for the
arrival process by inspection of the empirical distribution of
interarrival times. The interarrival times suffer from similar
censoring problems as the arrival times. Moreover, due to the
inhomogeneity of the proposed model, the interarrival times are not
well defined. For these reasons we find the use of interarrival time
statistics to be questionable. A more robust method could be to use the first few order statistics
for calibration and validation since these are very likely to stem
from strong and well separated signal components.   In either case,
when calibrating and validating multipath models based on estimation
of arrival times, the properties and biases caused by the delay
estimation procedure should be understood and factored in.

The model presented is constructed from
physical analysis based on mirror source theory. This approach
provides insight into how the environment (here the room) and the
system parameters (here the antennas and the transmitted signal) affect
the model. This insight is advantageous compared to an empirical
model. Empirical models, however, can be more easily fitted to
measurement data. It is therefore worth mentioning that the arrival
rate model considered here may be also used to motivate an empirical
model of the form
\begin{equation}
  \label{eq:38}
  \lambda(\tau) = \eta \cdot \tau^2 \cdot \mathds 1(\tau>0), 
\end{equation}
where the factor $\eta$ should be determined from measurements. The
complexity of this quadratic rate model is the same as the constant rate model, since
either are  defined by only a single parameter.  The 
model \ref{eq:38} is able to represent the specular-to-diffuse
transition observed experimentally for inroom channels. Such a
transition effect is not captured by the constant rate model.

\section{Conclusion}
\label{sec:conclusion}
We have proposed a stochastic model for the arrival times in an
in-room scenario. The proposed model is based on approximation of the
positions of mirror sources by spatial (3D) Poisson process. This
induces a non-homogenous Poisson process for the arrival times and a
model for the second moment of the power gain of a multipath component
conditioned on its arrival time. By construction, the path arrival
rate and power delay spectrum of the resulting stochastic multipath
model agrees with the mirror source model.  Nonetheless, the statistical
structure of the mirror source process and thus of the arrival times,
is not kept.

The proposed Poisson approximation is mathematically more convenient
than the mirror source model as it enables closed-form
derivation of expressions of a number of signal characteristics.  Here
we derive the cumulant generating functional for the received signal
and use it to obtain the kurtosis delay spectrum of the
received signal.  The kurtosis depends on the arrival rate in a very
direct fashion. In the high-bandwidth case, the arrival rate is
inversely proportional to the arrival rate. Due to the increasing
arrival rate, the pdf of the received signal depends on delay. At
small delays the received signal can differ significantly from a
Gaussian but as the delay increases, the pdf approaches a Gaussian.
Furthermore, we show that the order statistics of the arrival times,
i.e. the time of the $n$th arrival, follows a generalized gamma
distribution with the parameters determined by antenna coverage
fractions and the room volume. Based on the order statistics, we give
a closed form expression for the relative residual power after
removing the first $n$ arrivals.  Monte Carlo simulations show that
the proposed model agrees well to the mirror source model in terms of
power delay spectrum, kurtosis, order statistics of arrival times,
mean delay and rms delay spread.

The constant rate model, while having power delay spectrum
identical to the two other models, the does not predict well any of
the other studied characteristics (distributions of mean delay,  rms
delay spread and order statistics). Thus,  accurate
modelling the received signal using a stochastic multipath model
necessitate accurate  modelling of the arrival rate.  The constant
rate model as used in e.g. the Saleh-Valenzuela model is not able to
predict these characteristics






\appendices

\section{Generating Functionals}
\label{sec:appendix:gener-funct}
The characteristic functional for $y(t)$
evaluated for arbitrary probing function $\phi(t)$ is  defined as
 \cite{Rice1977,Snyder1991} 
\begin{align}
  \label{eq:18}
  C[\phi] &= \mathbb E\bigg[\exp\big(j\Re\int \phi(t) y(t)  dt\big )\bigg]
\end{align}
where $\Re$ denotes the real part. The complex natural logarithm of the characteristic functional is the cumulant
generating functional denoted by $K[\phi]$. By Kingmann's marking
theorem \cite{Kingman1993}, the marked point process  $\{(\tau_\ell,\alpha_\ell)\}$ with
forms a two-dimensional Poisson process with rate
$p(\alpha|\tau)\lambda(\tau)$. Then using Campbell's theorem
\cite{Kingman1993} and taking the logarithm we obtain
\begin{align}
  K[\phi]&= 
 \iint
  \big ( e^{j\Re\alpha\int \phi(t)s(t-\tau)dt} -1 \big)  p(\alpha|\tau) 
\lambda(t) d\alpha d\tau \\
&=\int 
  \bigg[ C_{\alpha|\tau}(\int \phi(t)s(t-\tau)dt)-1 
  \bigg] \lambda(\tau) d\tau 
\end{align}
where $C_{\alpha|\tau}(\cdot)$ is the characteristic
function for $p(\alpha|\tau)$.

The probing function plays the same role as the variable introduced in 
the more widespread characteristic and cumulant generating functions. Evaluating the
cumulant generating functional for $\phi(t) = \nu\delta(t)$,  we obtain the cumulant 
generating function for $y(t)$ for any given time $t$:
\begin{equation}
  \label{eq:95}
K(\nu)=\int 
  \bigg[ C_{\alpha|\tau}(\nu s(t-\tau))-1 
  \bigg] \lambda(\tau) d\tau   
\end{equation}
Cumulants of $y(t)$ can now be computed by complex
differentiation as
\begin{align}
  \label{eq:87}
\kappa_{m:n}(t)  &= 
  \frac{\partial^{m+n}}{ \partial^m \nu \partial{ \nu^*}^{n}}
  K(\nu)\bigg |_{\nu = 0}\\
&=  \int s(t-\tau)^{m} s(t-\tau)^{*n}
 \mathbb E[  \alpha^{m}{\alpha^*}^{n}| \tau ] \lambda(\tau) d \tau. 
\end{align}
Considering the gains to be circular random variables, the moments $\mathbb E[
\alpha^{m}{\alpha^*}^{n}| \tau ]$ are zero for   $m\neq n$ and all
odd cumulants (and moments) of $y(t)$ vanish. The even cumulants in
\eqref{eq:96} are obtained with $m=n$.   In particular for $m=n=1$ we obtain
the delay power spectrum, i.e $\kappa_{1:1}(\tau) = P(\tau)$.


\section{Kurtosis Estimation for Complex Circular Variables}
\label{sec:KurtEstAppendix}
Here derive an estimator for the fourth cumulant of a circular complex random
variable $X$ based iid.\ observations $X_1,\dots,X_N $.  For a circular
complex variable, the fourth cumulant, fourth and second moments are related as \cite{Schreier2010}
\begin{equation}
  \label{eq:24}
  \kappa_{2:2}[X] = \mathbb E[|X|^4] - 2 \mathbb E[|X|^2]^2. 
\end{equation}
We seek an estimator of the form 
\begin{equation}
\label{eq:36}
  \hat\kappa_{2:2}[X] = c_1 \sum_{n=1}^N {|X_n|^4} - c_2 \left(\sum_{n=1}^N |X_i|^2\right)^2\!.
\end{equation}
For an unbiased estimator,    $\mathbb E[  \hat\kappa_{2:2}[X]] =
\kappa_{2:2}[X]$. By  using \eqref{eq:24} and some
straight-forward manipulations we obtain 
\begin{equation}
  \label{eq:25}
  c_1 = \frac{N+1}{N(N-1)}, \quad \text {and }  \quad  c_2 =
  \frac{2}{N(N-1)}, \quad N>1.
\end{equation}
Note this estimator differs from the unbiased estimator
obtained for real valued data derived in
\cite{Blagouchine2009}.  The kurtosis is then estimated as
$\hat\kappa_{2:2}[X]/   \hat\kappa_{1:1}^2[X]$.

\section{Simulation of Inhomogeneous Poisson Process}
\label{sec:SimAppendix}

The inhomogeneous Poisson point process $\mathcal
T_{\mathrm {PPP}}$   can be simulated on a finite interval $ [0,\tau_{\max}]$ by a two-step
procedure: 1) Draw
a Poisson count $N(\tau_{\mathrm {max}})$ with mean $  \mathbb E [N(\tau_{\max})] $ as specified by  the
model. 2) Draw  $N(\tau_{\mathrm {max}})$  iid.\ delay variables according to the
pdf in  \eqref{eq:30}.
The corresponding cdf 
\begin{equation}
  \label{eq:27}
  F(\tau) = 
  \begin{cases}
0 & \tau<0\\
\frac{\tau^3}{\tau_{\max}^3} &  0\leq\tau
\leq \tau_{\max}\\    
1 & \tau<\tau_{\max}.
  \end{cases}
\end{equation}
is easy to invert, and therefore we can use the inverse cumulative
distribution method \cite{Davison2003}. This amounts
to transforming a variable $U$ uniformly  distributed on the interval
$[0,1]$ according to $  \sqrt[3]{U} \tau_{\max}$. 


\bibliographystyle{IEEEtran}
\bibliography{/Users/troels/Documents/References/referencedatabase}
\end{document}